\documentclass[twocolumn,showpacs,preprintnumbers,amsmath,amssymb]{revtex4}

\usepackage[unicode]{hyperref}
\usepackage[caption=false]{subfig}
\usepackage{graphicx}
\graphicspath{ {./images/} }
\usepackage{dcolumn}
\usepackage{bm}
\usepackage[caption=false]{subfig}
\usepackage{graphicx,epsf}
\begin{document}

\title{Dynamic scaling and stochastic fractal in nucleation and growth processes}

\author{Amit Lahiri$^{\dagger}$, Md. Kamrul Hassan$^{\dagger}$, Bernd Blasius$^\$ $
and J\"urgen Kurths$^\ddagger$}
\affiliation{
$^\dagger$ University of Dhaka, Department of Physics, Theoretical Physics Division, Dhaka 1000, Bangladesh \\ 
$^\$ $ Institute for Chemistry and Biology of the Marine Environment (ICBM)
Carl-von-Ossietzky University Oldenburg PF 2503, 26131 Oldenburg, Germany \\
$^\ddagger$ Potsdam Institute for Climate Impact Research (PIK), Postfach 601203, 14412 Potsdam, Germany 
}

\begin{abstract}

A class of nucleation and growth models of a stable phase
(S-phase) is investigated for various different growth velocities. It is shown that for 
growth velocities $v\sim s(t)/t$ and  $v\sim x/\tau(x)$, 
where $s(t)$ and $\tau$ are the mean domain size of the metastable phase (M-phase) and the mean
nucleation time respectively, the M-phase decays following a power law. Furthermore, snapshots at different
time $t$ are taken to collect data for the distribution function $c(x,t)$ of the domain size $x$ of M-phase
are found to obey dynamic scaling. Using the idea of data-collapse we show that each 
snapshot is a self-similar fractal. However, for  
$v={\rm const.}$ like in the classical Kolmogorov-Johnson-Mehl-Avrami (KJMA) model
and for $v\sim 1/t$ the decay of the M-phase are exponential and they are not accompanied 
by dynamic scaling. We find a perfect agreement between numerical simulation and analytical results.
\end{abstract}


\maketitle

{\bf  A class of nucleation and growth process is studied analytically by solving an integro-partial differential equation
and verified numerically by Monte Carlo simulation. The growth velocity is defined as the ratio of the distance traveled $s$ 
by the new phase and the magnitude of time $t$ is needed to travel that distance. We first
choose constant growth velocity by assuming both $s$ and $t$ equal to constant and reproduce the results of
the much studied classical Kolmogorov-Johnson-Mehl-Avrami (KJMA) model. Choosing one of them a constant still makes the exponential
decay of the meta-stable phase like the KJMA model. However, the growth velocity is so chosen that neither $s$ nor $t$ is constant
we find that the meta-stable phase decays following a power-law. Such power-law is also accompanied by the emergence of fractal. 
The self-similar property of fractal is verified by showing that the system exhibits dynamic scaling revealing a self-similar
symmetry along the continuous time axis. According to Noether's theorem there must exist a conserved quantity and
we do find that the $d_f$th moment, where $d_f$ is the fractal dimension, is always conserved.}

\section{Introduction}

The formation of any phase is usually a process that starts first  by
nucleation of new phase on old phase, which is called metastable phase, 
followed by growth of the new phase which eventually becomes a stable phase
(S-phase). This mechanism of nucleation and growth represents one of the most fundamental 
topics of interest in both science and technology \cite{kn.christian}. It plays a key role in metallurgical
applications as well as in many seemingly unrelated fields of research. Phase separation and
coarsening \cite{ref.bray}, electrodeposition of metals at electrodes via nucleation and growth \cite{ref.kong},  melting of stable glasses
\cite{ref.jack},  dendritic growth \cite{ref.langer}, kinetics of crystal growth \cite{kn.crystallization}, the domain switching
phenomena in  ferroelectrics \cite{kn.ferroelectrics},  the spread of ecological 
invaders \cite{ref.arim} and growth of breath figures \cite{ref.meakin} are just a few examples of such systems. Besides, G. Giacomelli et al.
studied an interesting case in which nucleation occurs in time not in space \cite{ref.giacomelli}. 
The nature of phase transition, which is governed by nucleation and 
growth of domains of S-phase, is well-known as first order phase transition. One of the characteristic
features is that during the transition stable and metastable regions
coexist at the same time. 
Much of our theoretical understanding of such phenomena is provided by the
Kolmogorov-Johnson-Mehl-Avrami (KJMA) model, which has been formulated
independently by Kolmogorov, Johnson and Mehl and Avrami in and around the
$1940$s \cite{kn.kolmogorov}.  This model still remains one of the most studied
theories of nucleation and growth processes as it is often the only means of 
interpreting the experimental data that help gain insights into the process.

The KJMA theory is valid under the assumptions that (a) nucleation events are Poissonian in nature, 
(b) seeds grow with constant velocity while keeping fixed geometrical shape and orientation, and (c) the 
system is homogeneous in space and time. In the context of nucleation and growth phenomena, one
of the central quantities of interest is 
the fraction of the M-phase $\Phi(t)$ that still survives at time $t$.
According to the KJMA theory, this quantity in $d$ dimensions follows an
exponential decay known as the Kolmogorov-Avrami law
\begin{equation}
\label{eq:k-a}
\Phi(t)=\exp\Big [-{{\Omega_d}\over{d+1}}\Gamma v_s^d t^a\Big ],
\end{equation}
where, the Avrami exponent $a=d+1$ \cite{kn.kolmogorov}, $\Gamma$ describes the constant 
nucleation rate per unit volume and $\Omega_d$ is the constant volume factor of the $d$ 
dimensional hypersphere e.g., $\Omega_d =1,\pi,4\pi/3$  for $ d=1,2,3$ respectively.

The derivation of  correlation functions and their connection to the
scattering cross-section \cite{kn.sekimoto,kn.sekimoto_ng}, the theory of grain-size
populations \cite{kn.population}, and the generalization of the KJMA theory to
multiple stable phases \cite{kn.multiphase} etc. have played a significant role
in understanding the phenomena. Besides, these studies have provided a better 
means of interpreting the experimental data. On the
other hand, there have been reports that the experimental data in  some cases do
not fit a straight line in the plot of $\log(-\log[\Phi(t)])$ against $\log[t]$ revealing that it violates
the Kolmogorov-Avrami law \cite{kn.price,kn.sessa}. Recently, it has also been observed that nucleation and growth 
processes results in the emergence of fractal \cite{kn.fractal_nucleation_1,
kn.fractal_nucleation_2,kn.fractal_nucleation_3, kn.fractal_tieng, ref.riedel}. 
Obviously, constant growth velocity does not result in fractal. However, the
nature of growth mechanism responsible for the emergence of fractal is not yet fully understood. 
This observation clearly raises some concerns and
hence it requires further theoretical interest in order to find variants of the 
model which would be suitable under different physical situations. This is exactly the purpose of the present work.

In this article, we investigate a class of Kolmogorov-Johnson-Mehl-Avrami nucleation and 
growth model in one dimension for four different choices of the growth velocities.
In particular, we choose (A) $v=const.$, (B) $v=\sigma/t$, (C) $v=ks(t)/t$ 
and (D) $v=mx/\tau(x)$ where $\sigma,k$ and $m$ are constants, $s(t)$ is the mean domain size and $\tau(x)$ is 
the mean nucleation time \cite{kn.maslov}. The idea to move away from the constant growth velocity is born by the 
observation that a growth velocity which may change in the course of time and space. Such choices
may provides a more realistic description and hence will become more suitable under various natural conditions than constant growth velocity. 
Usually, as the time proceeds, size and shape of both phases 
change continuously, altering the condition that ultimately determines the growth velocity. 
We solve the corresponding rate equation for each choice of the
growth velocity and obtain the exact solution for the domain size distribution function
$c(x,t)$ of M-phase. Alongside, we also give exact algorithm for each case to solve them by numerical simulation.
Using the idea of data collapse we show that the analytical results are in perfect agreement with our
numerical simulations. It has been found that the M-phase decays exponentially with time 
for models (A) and (B) whereas in the case of models (C) and (D), it decays following a power-law. Moreover, 
the domain size distribution function $c(x,t)$ of models (C) and (D) is found to obey dynamic 
scaling $c(x,t)\sim t^{\theta z}\phi(x/t^z)$ revealing that it evolves with time in a self-similar 
fashion and eventually emerges as fractal. We find the kinetic exponent $z=1$ and the mass exponent $\theta=1+d_f$.
Their numerical values are fixed by the dimensional consistency and the conservation principle.

The rest of this article is organized as follows. Section~\ref{sec-formulation} formulates 
the  rate equation for nucleation and growth model in one dimension. We also attempt
to solve the rate equation for generalized growth velocity. 
In Section~\ref{sec-model_a}-\ref{sec-model_d}, we solve the rate equation
for four different growth velocities and discuss the corresponding scaling theory.
For each choice of the growth velocity exact algorithm is given and extensive Monte Carlo simulation is 
performed to corroborate all the analytical results namely decay of the M-phase, fractal dimension, a 
conserved moment and dynamic scaling. In Section~\ref{sec-conc}, we make concluding remarks and leave some open questions.

\section{Formulation of the Rate equation} \label{sec-formulation}

The right choice of the growth velocity should depend on the detailed nature of
the system under investigation. In the present work, we make a few simple choices
so that we can handle the problem analytically.
Further, we took extra care on the dimensional
consistency with the terms describing the nucleation mechanism.   
Our aim is to learn how much
effect the growth velocity alone has on the dynamics of the system. To obtain a mathematical formalism
of the phenomena, we find it useful to treat nucleation and growth mechanisms
separately and derive their respective governing equations. To do it, we define
the distribution function $c(x,t)$ as the concentration of domain size $x$ of M-phase
at time $t$. First, we derive the rate equation for the nucleation
mechanism. The random nucleation of seeds of S-phase can be thought of as a
random sequential adsorption (RSA) of monodisperse, size-less particles on a
substrate of M-phase. The distribution function  $c(x,t)$ then obeys the
following well known rate equation for RSA \cite{kn.sekimoto_ng,kn.naim}
\begin{equation}
\label{eq:nucleation}
\left. {{\partial c(x,t)}\over{\partial t}} \right |_{{\rm nucleation}} =  -x\, c(x,t)+2\int_x^\infty c(y,t)dy .
\end{equation}
%
The two terms on the right hand side of Eq. (\ref{eq:nucleation}) represent the
decrease and increase of M-phase intervals of size $x$ due to the nucleation
of point-like seeds of S-phase at locations $x$, and $y>x$, respectively.  The
factor $2$ in the integral term takes into account that every nucleation event
creates two new domains of M-phase either of which can be of size $x$. 

To derive an expression for the growth mechanism, we consider that
once  a seed of S-phase has nucleated at some point on M-phase, it keeps growing
with some velocity $v(x,t)$.  Here we assume a simplified generalized case where the growth velocity
may depend on the size $x$ of the M-phase and the time $t$.  
In general, the growth velocity may also depend on the size of the S-phase
segment, but that goes beyond the scope of the present work. 
To derive the growth equations, note that the growth of the S-phase occurs at
the expense of the decay of the M-phase. That is, the  size of the M-phase on
either side of the growing seeds are shrinking with the same velocity $v$ with 
which seeds of the S-phase are growing. 
The growth term can be deduced by taking into account all the possible ways
in which the distribution function $c(x,t)$ remains in the domain size
range $[x,x+dx]$ after a span of infinitesimal time $dt$
\begin{eqnarray}
c(x,t+dt)  & = & \big [1-2 v(x,t)dt/dx \big ]c(x,t) + \\ \nonumber &  &  
\big [2 v(x+dx,t)dt/dx \big ]c(x+dx,t).
\end{eqnarray}
Here, $v(x,t)dt/dx$ is the fraction of concentration $c(x,t)$ that is lost in
time $dt$  due to the growth of the S-phase with velocity $v(x,t)$, and the
factor 2 takes into account of the fact that M-phase segments are reduced from both sides. 
Thus
\begin{equation}
\label{eq:growth}
\left. {{\partial c(x,t)}\over{\partial t}}\right |_{{\rm growth}}
= 2 {{\partial}\over{\partial x}}\big[v(x,t)c(x,t)\big]
\end{equation}
Putting the nucleation and growth terms together, the general rate equation for the random nucleation followed by space-time dependent continuous  growth therefore is
\begin{eqnarray}
\label{eq:RateEq}
{{\partial c(x,t)}\over{\partial t}}& = & -xc(x,t)+2\int_x^\infty c(y,t)dy \\ \nonumber & + & 2 {{\partial}\over{\partial x}}\big[v(x,t)c(x,t)\big].
\end{eqnarray}
We now attempt to solve Eq. (\ref{eq:RateEq}) subject to the initial conditions 
\begin{equation}
\label{eq:initial}
c(x,0)={{1}\over{L}}\delta(x-L), \hspace{0.6cm} 
\lim_{L \longrightarrow \infty}\int_0^\infty c(x,0)dx =0.
\end{equation}
This ensures that there is no seed of the S-phase at $t=0$.  Once we know
$c(x,t)$, we can immediately find the fraction of  the M-phase that remained
un-transformed 
\begin{equation}
\Phi(t)=\int_0^\infty xc(x,t)dx
\end{equation}
and the number  density of the domain size of the M-phase 
\begin{equation}
N(t)=\int_0^\infty c(x,t)dx.
\end{equation}
The fraction of  the M-phase covered by the S-phase is  related to $\Phi(t)$ via
$\theta(t)=1-\Phi(t)$ that evolves with time as 
\begin{equation}
    {{d\theta(t)}\over{dt}}=\int_0^\infty v(x,t)c(x,t)dx.
\end{equation} 
The quantities $\Phi(t)$ and $N(t)$ can also  be used to obtain
the time dependence of the average interval size $s(t)=\Phi(t)/N(t)$.


In the context of the present work, it is the choice of the growth velocity $v(x,t)$ that will 
define different models and accordingly we study the following four different choices:
(A) $v(x,t)=v_0$, the traditional constant growth velocity (classical KJMA model), (B) $v(x,t)=\sigma/t$ where 
$\sigma$ is a constant and bears the dimension of length, (C) $v(x,t)=s(t)/t$ where $s(t)$ is the mean interval
size of the M-phase at time $t$ and finally (D) $v(x,t)=m x/\tau(x)$ where $\tau(x)$ is the mean nucleation 
time and $m$ is a dimensionless positive constant \cite{kn.maslov}.

For growth velocities $v=v(t)$ which are  independent of the variable $x$ (as in
our models {\it (A)}\ -  {\it (C)}), the general solution of Eq. \ (\ref{eq:RateEq}) can
be obtained by substituting the ansatz
\begin{equation}
\label{eq:3}
c(x,t)=A(t)\exp[-xt].
\end{equation}
Here, the time dependent pre-factor $A(t)$ obeys the following ordinary differential equation 
\begin{equation}
\label{eq:4}
{{d\ln A(t)}\over{dt}}=2/t-2v(t)t.
\end{equation}
We have to solve Eq. (\ref{eq:4}) subject to the initial condition $A(0)=0$,
followed by  Eq.  (\ref{eq:initial}). To proceed further and for clarity, we
will treat for every choice of growth velocity separately and independently.

First, we note that the time dependence of $s(t)$ is independent of $A(t)$
\begin{equation}
s(t)=\frac{\Phi(t)}{N(t)} = 
\frac{ A(t) \int_0^\infty x \exp[-xt] \, dx}{
A(t) \int_0^\infty \exp[-xt]\, dx}.
\end{equation}
Solving the integrals we find that in this case of $v=v(t)$ the mean interval size of M-phase shrinks as 
\begin{equation}
s(t)= 1/t .
\label{eq:st}
\end{equation}
In choosing the growth velocity, we shall make use of this relation.
Interestingly, the same expression also holds for a model without the growth term which is known as random scission model \cite{kn.ziff}.

\section{Model A}\label{sec-model_a}

\begin{figure}

\centering

\subfloat[]
{
\includegraphics[height=4.0 cm, width=4.2 cm, clip=true]
{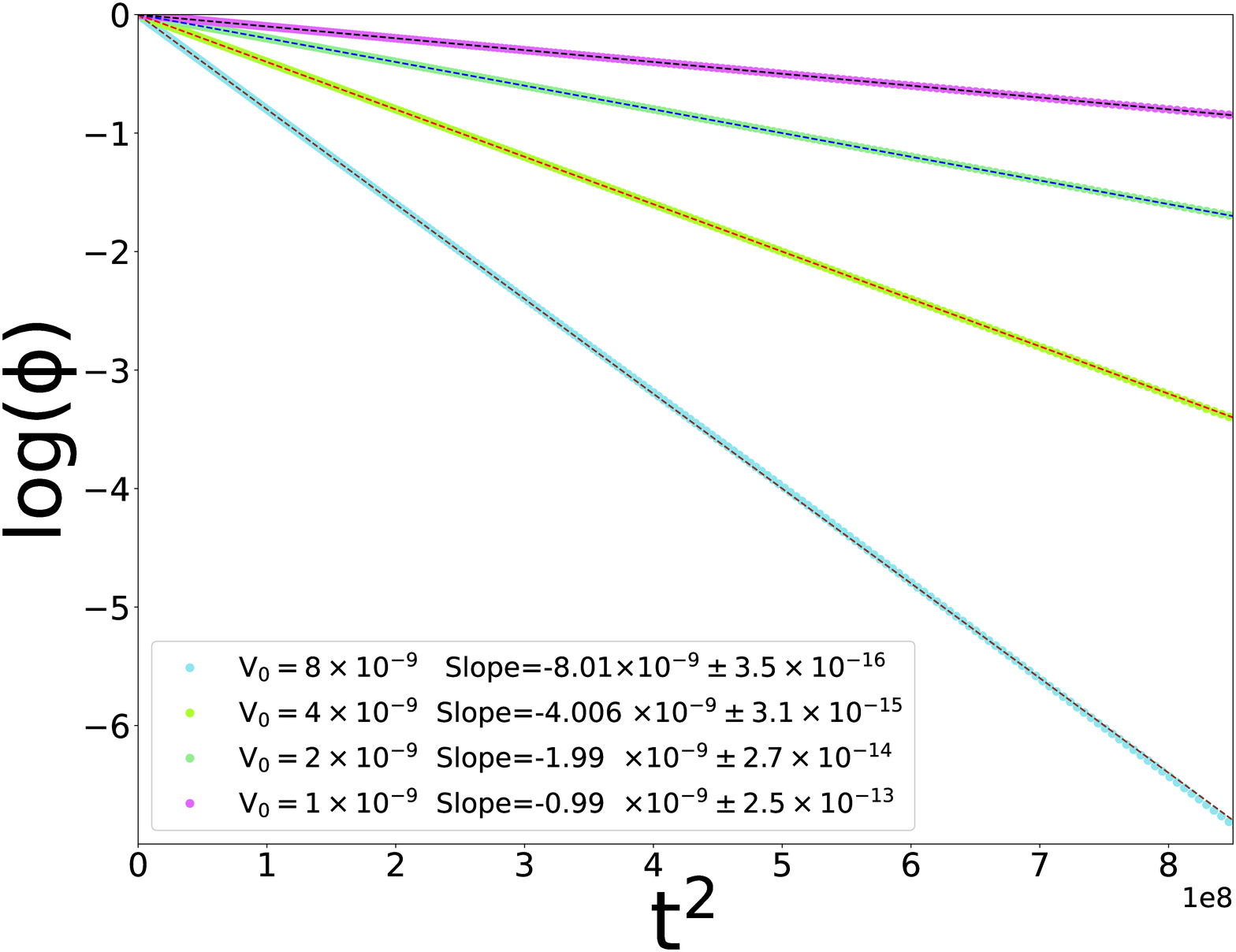}
\label{fig:1a}
}
\subfloat[]
{
\includegraphics[height=4.0 cm, width=4.2 cm, clip=true]
{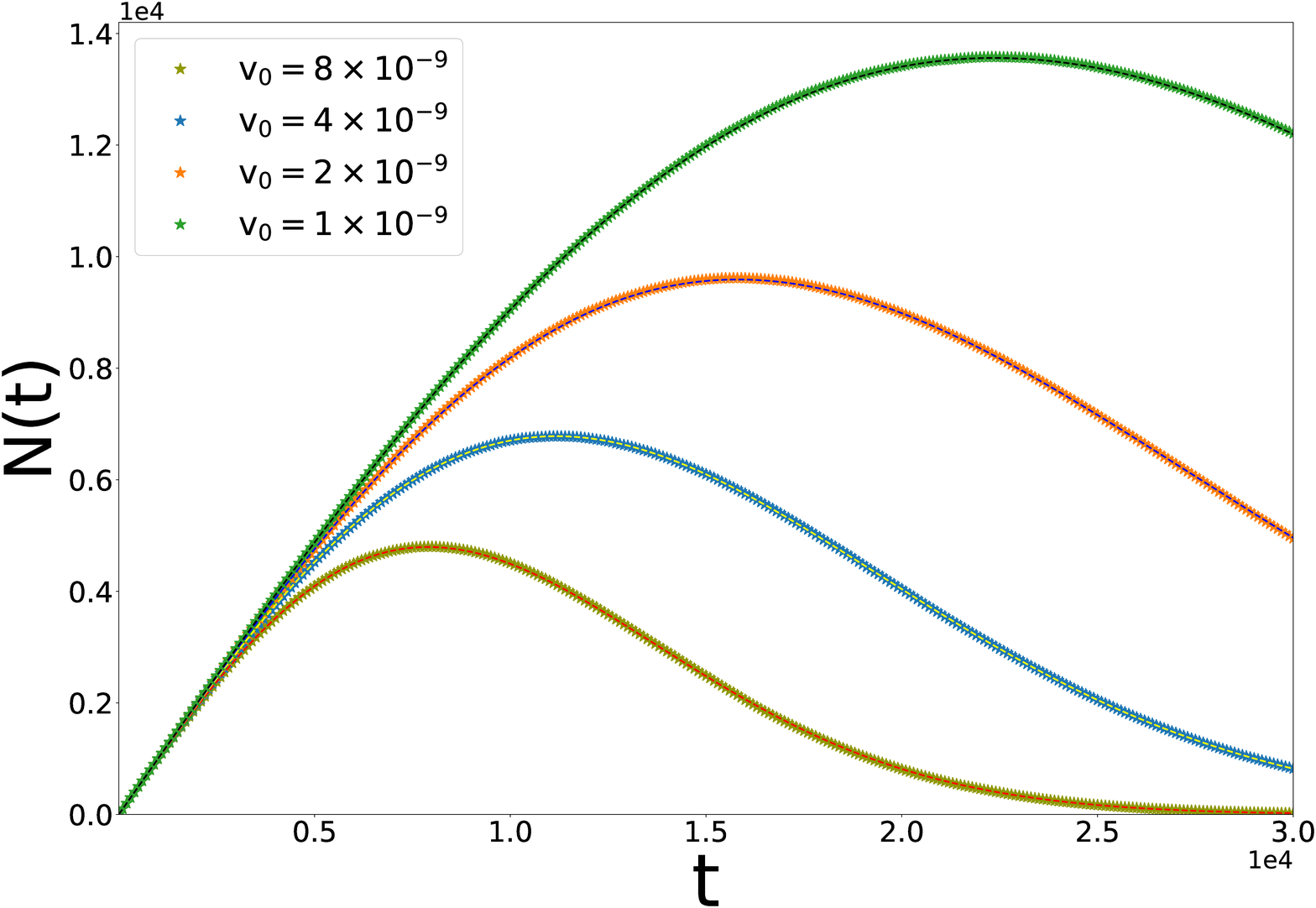}
\label{fig:1b}
}

\caption{ (a) We plot $\log(\phi(t))$ versus $t^2$ for different constant growth velocities to see how the $M$-phase $\phi(t)$ decays 
with time. (b) Plots of $N(t)$ versus $t$ is shown for different constant growth velocities. 
} 

\label{fig:1ab}
\end{figure}

We first consider the classical KJMA model defined by the constant velocity of domain of S-phase 
$v(t)=v_0$. The goal is to reproduce the known results so that they help build confidence 
on the rate equation approach and help appreciate its simplicity. 
Solving the problem rest on finding the distribution function $c(x,t)$ which means coding all the essential 
information regarding this model. After substitution of $v(t)=v_0$ into Eq. (\ref{eq:4}) and a straightforward integration we immediately obtain 
$A(t)=t^2\exp[-v_0t^2]$ and the solution of Eq. (\ref{eq:RateEq}) for the classical KJMA model is given by
\begin{equation}
 \label{eq:5}
c(x,t)=t^2\exp[-xt-v_0t^2].
\end{equation} 
Using this in the definition of $\Phi(t)$, we immediately find that
\begin{equation}
\label{eq:phi_a}
    \Phi(t)=\exp[-v_0t^2].
\end{equation} 
It implies that the plots of $\log(\Phi(t))$ versus $t^2$
should give a straight line with slope equal to $-v_0$. This is indeed the case
as shown in Fig. (\ref{fig:1a}). Note that $v_s=2v_0$ where $v_0$ is the velocity of front of the S-phase that 
moves forward transforming the M-phase into S-phase within it. Thus, the velocity with which each seed grows is $v_s=2v_0$ 
and substituting it into the expression for $\Phi(t)$, we can immediately recover the celebrated Kolmogorov-Avrami law 
 in one dimension as given in Eq.(\ref{eq:k-a}), except the factor $\Gamma$. Note that Eq. (\ref{eq:nucleation}) describes 
 the random  sequential nucleation of one seed at
each time step and hence $\Gamma=1$ in the context of the present rate equation approach. 
The rate equation approach thus clearly demonstrates its simplicity and brevity. This is however true only for one dimension. 
In the context of nucleation and growth phenomena, the 1d model can in fact fully capture the qualitative behavior of the key 
quantities of interest, namely the decay of the M-phase. The Kolmogorov-Avrami law itself is a testament to its justification 
which clearly shows that the exponential decay of the M-phase is common to all dimensions.

Integration of Eq. (\ref{eq:5}) gives the number density of domain of M-phase 
$N(t)= te^{-v_0t^2}$. This expression reveals that  during the very early
stages $N(t)$ rises linearly due to the nucleation of  the S-phase, whereas at
the late stages $N(t)$ decreases exponentially, which reflects a fast
coalescence of the neighboring stable phases as seen in Fig. (\ref{fig:1b}) \cite{kn.naim}. 
By taking the ratio  $\Phi(t)/N(t)$ we confirm that the mean domain size
$s(t)$ of the M-phase decays as  $s(t)= t^{-1}$.  
To verify these results we have done Monte Carlo simulation based on the following elaborate 
algorithm. One time unit of the process in one dimension can be defined as
follows:  

\begin{itemize}
\item[{(\it i)}] At the $j$th step say there are $n$ number of domains of M-phase of size
$x_1,x_2,...,x_n$.
\item[{(\it ii)}] Generate a random number $R$ with a uniform distribution within the unit 
interval $[0,1]$. 
\begin{description}
\item[(\emph{a})] Check which of the domains of M-phase contain $R$. Say, it is the
$k$th domain whose size is $x$. Then sow a point-like seed exactly at $xR$ and go to step (iii).
\item[(\emph{b})] If $R$ falls within S-phase then increase the iteration step by one unit, and 
go back to step $(ii)$.
\end{description}
\item[{(\it iii)}] Increase the size of the seed of S-phase on either side by the constant value $v_0$
independently. If the domain size of M-phase in any side is less than $v_0$ then the growth of
the S-phase ceases immediately at the point of contact, while it is continuing elsewhere. 
\item[{(\it iv)}] Increase all the existing domains of S-phase in the same way as described in step (iii).
\item[{(\it v)}] Increase the iteration step and the number of domains of M-phase by one unit. 
\item[{(\it vi)}] Go to step (ii) and continue the process {\it ad infinitum}. 
\end{itemize} 
In Fig. (\ref{fig:1ab})  we show the plots of our various results and find that both 
analytical solutions of the KJMA-process are in perfect 
agreement to direct numerical simulations.

\section{Model B}\label{sec-model_b}

\begin{figure}

\centering

\subfloat[]
{
\includegraphics[height=4.0 cm, width=4.2 cm, clip=true]
{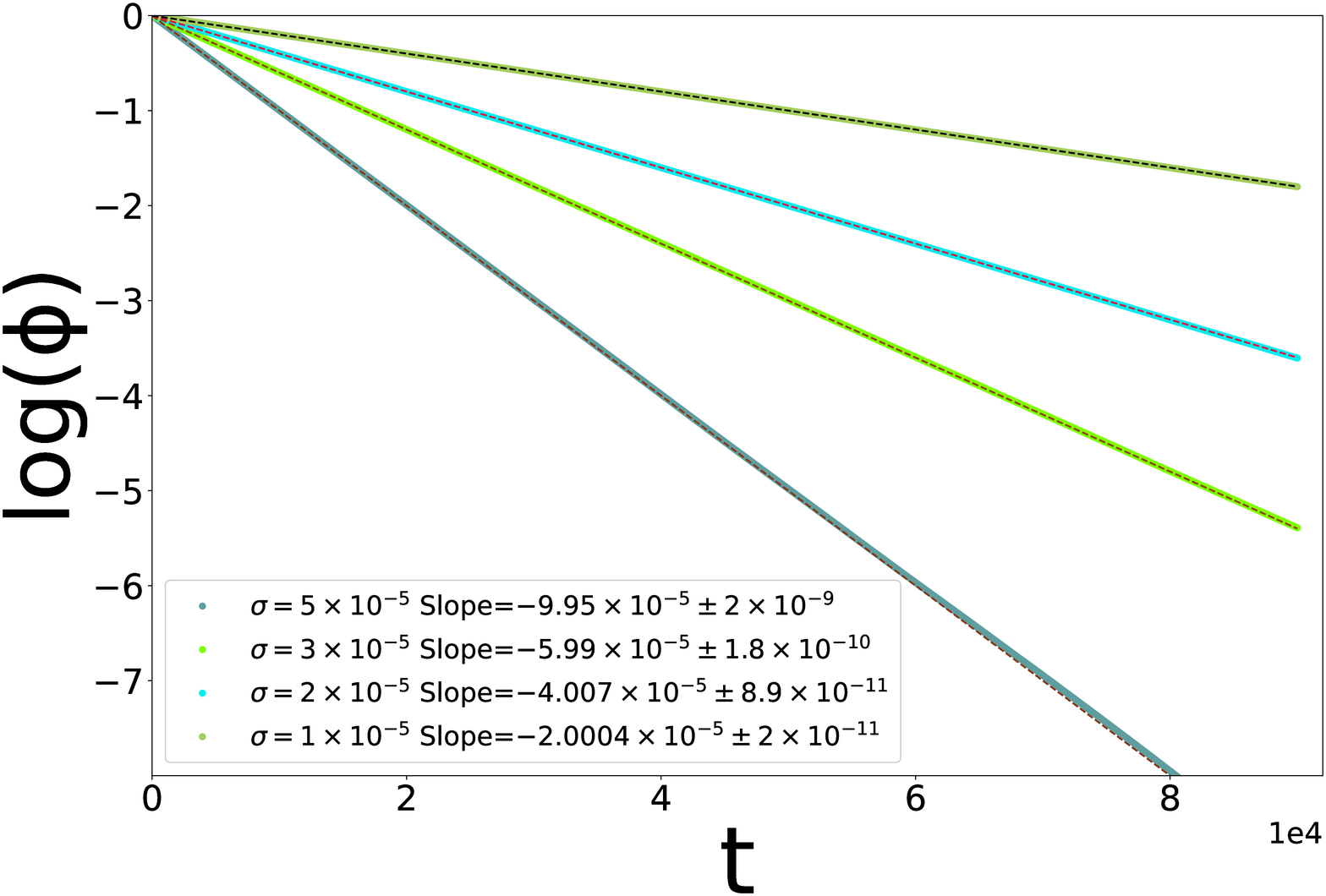}
\label{fig:2a}
}
\subfloat[]
{
\includegraphics[height=4.0 cm, width=4.2 cm, clip=true]
{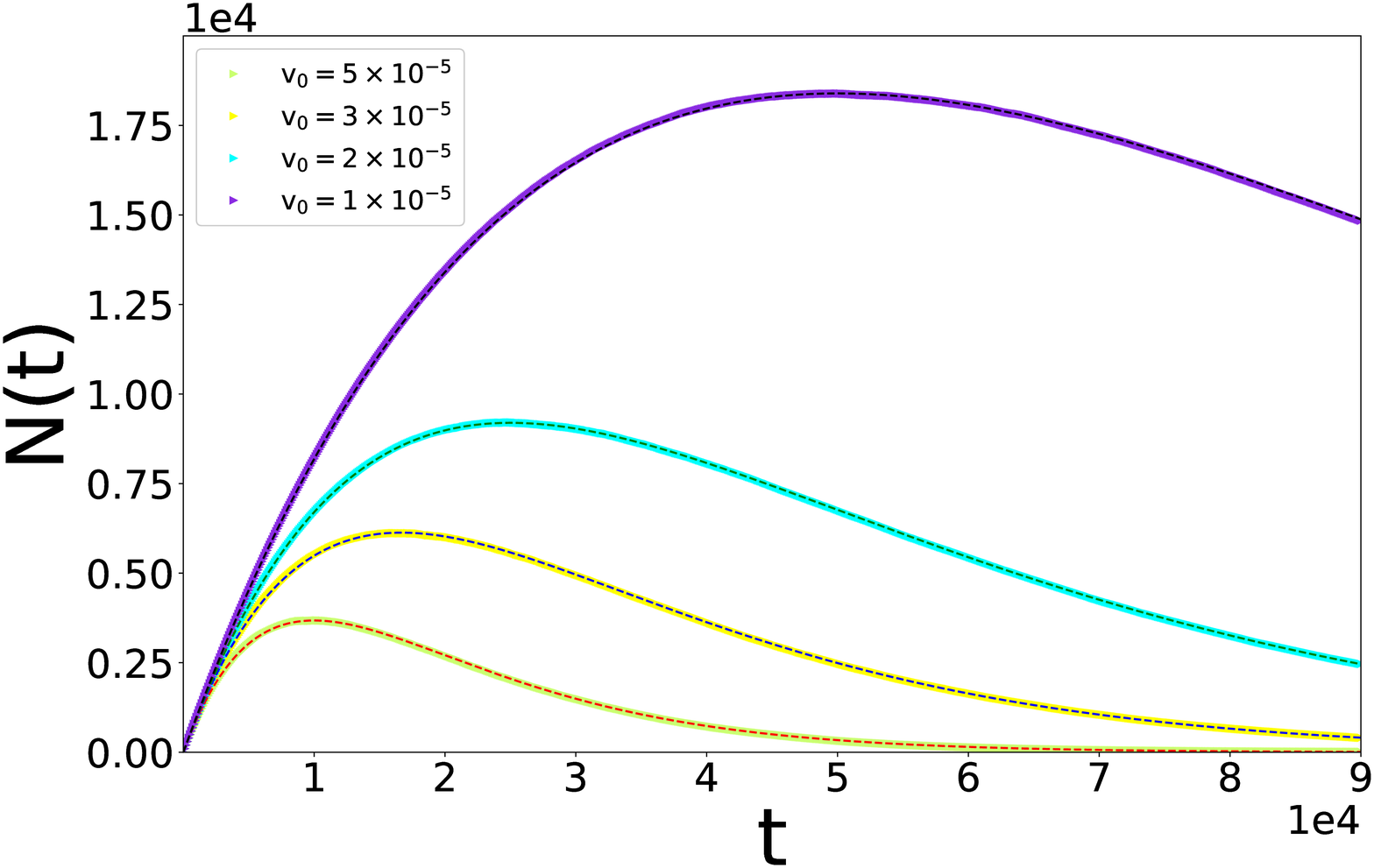}
\label{fig:2b}
}

\caption{ (a) Plots
of $\log(\phi(t))$ versus $t$ of model B are drawn for different $\sigma$ value that determine
the extent of growth velocities $v$ since $v=\sigma/t$. We find straight lines with slope always 
equal to $-2\sigma$ which is in complete agreement with our analytical solution given by Eq. (\ref{eq:phi_b}). 
(b) The number of domains $N(t)$ of $M$-phase versus $t$ are shown for model B as a function of time 
for different constant growth velocities.
} 

\label{fig:2ab}
\end{figure}

We now solve Eq. (\ref{eq:4}) for the growth velocity $v(t)=\sigma/t$
to give
\begin{equation}
 \label{eq:6}
c(x,t)=t^2\exp[-(x+2\sigma)t].
\end{equation} 
In this case too, we find that the fraction of  the M-phase decays
exponentially 
\begin{equation}
\label{eq:phi_b}
    \Phi(t)=\exp[-2 \sigma t],
\end{equation} 
but slower than that for constant 
velocity. Here the Avrami exponent corresponds to
the one that typically known for heterogeneous nucleation and growth processes
 despite the fact in the present case it strictly
describes the homogeneous nucleation \cite{kn.heterogeneous1}. In Fig.
(\ref{fig:2a}) we show plots of $\log(\Phi(t))$ versus $t$ and find a set of straight lines with slope
always equal to $2\sigma$ which are in perfect agreement with Eq. (\ref{eq:phi_b}).
This proves that the Avrami exponent not
only depends on the nature of the nucleation process but also on the exact choice of 
the growth velocity. In a similar way we find that the number density of  the M-phase varies as 
\begin{equation}
\label{eq:number_b}
    N(t)= te^{-2\sigma t}.
\end{equation} 
This reveals that at the early  stage $N(t)$ rises linearly like the model {\it (A)}. However, at the late stage, coalescence events take place less frequently than in the model {\it
(A)} as shown in Fig. (\ref{fig:2b}). On the other hand, $N(t)/t$ varies following the same relation with time as $\Phi(t)$ given 
by Eq. (\ref{eq:phi_b}). Also for model (B) analytical and numerical results show perfect agreement.

To verify these results we have done Monte Carlo simulation based on the
algorithm described in Model A except step (iii) which is replaced as follows. 
\begin{itemize}
\item Increase the size of the seed of S-phase on either side by $\sigma/j$
independently provided domain of S-phase is greater than $\sigma$ in both sides. However,
if the domain size of S-phase in any side is less than $\sigma/j$ then the growth ceases immediately at the point of contact, while it continues elsewhere. 
\end{itemize}
Plots in Fig. (\ref{fig:2ab}) are drawn using numerical data based on the above algorithm. We find that our analytical solutions
and numerical simulations are in perfect agreement.

\section{Model C}\label{sec-model_c}

\begin{figure}

\centering

\subfloat[]
{
\includegraphics[height=4.0 cm, width=4.2 cm, clip=true]
{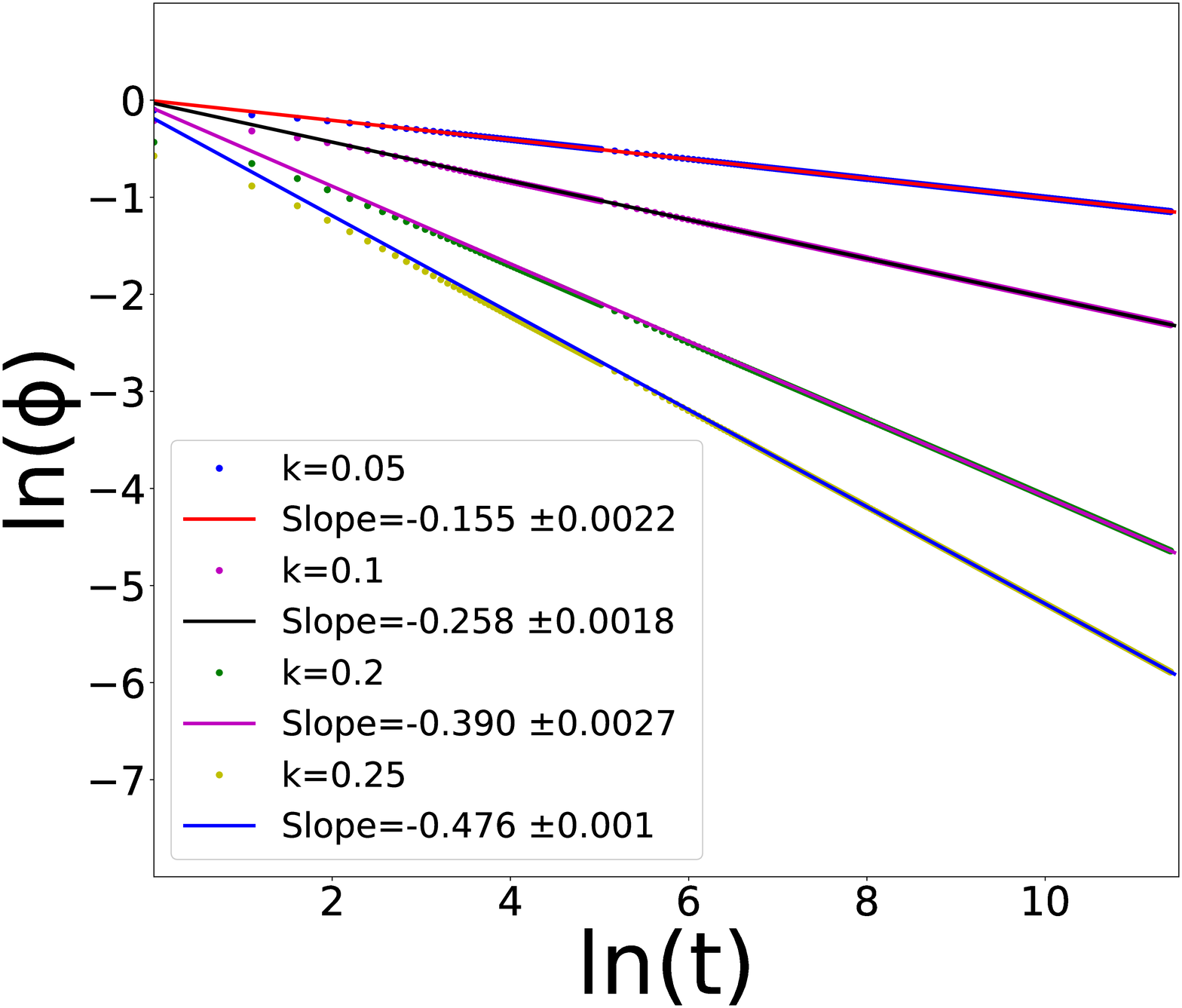}
\label{fig:3a}
}
\subfloat[]
{
\includegraphics[height=4.0 cm, width=4.2 cm, clip=true]
{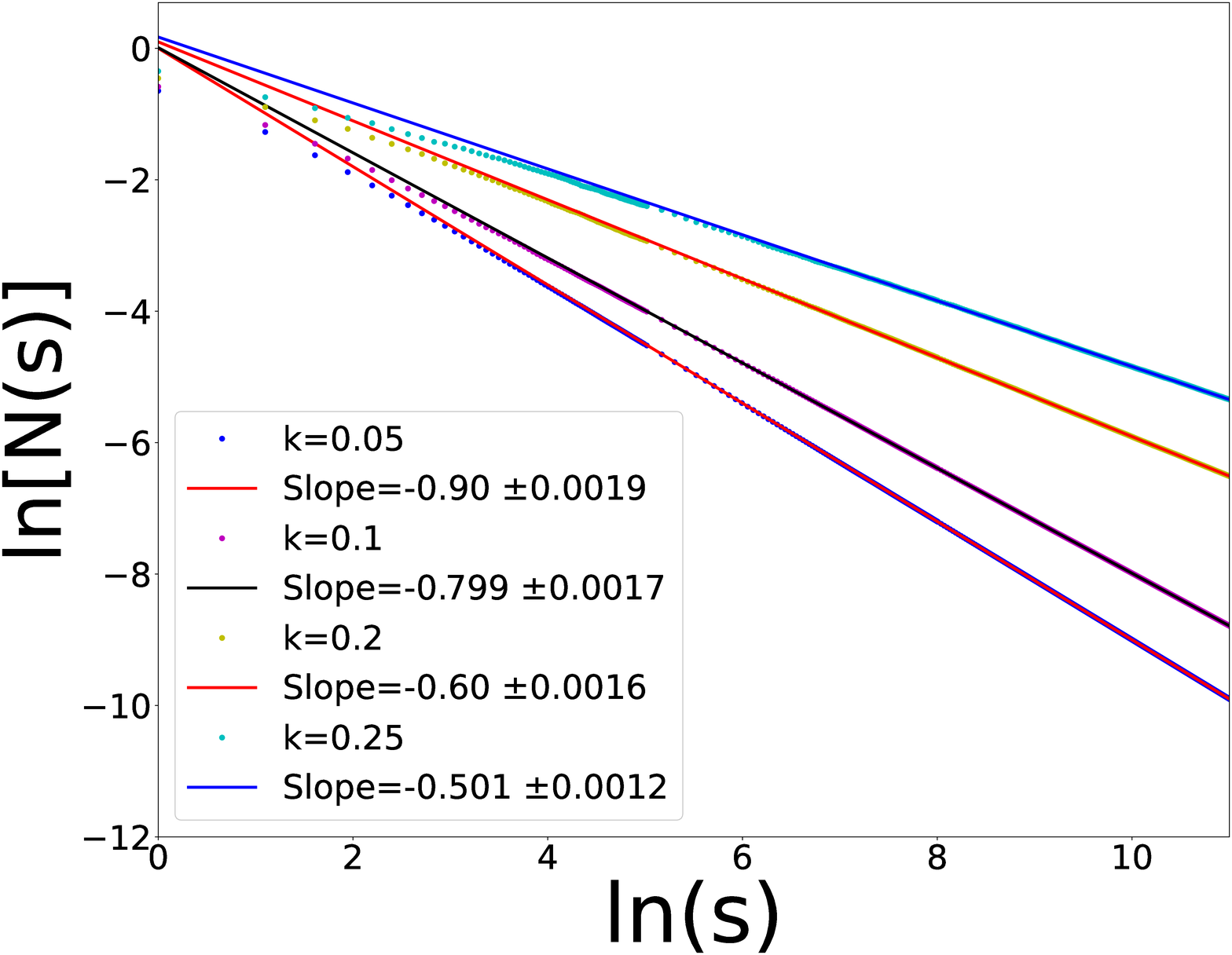}
\label{fig:3b}
}

\caption{ (a) Plots of $\log(\Phi(t))$ versus
$\log(t)$ are drawn for different $k$ values. The result is a set of straight lines with slopes
equal to $-2k$ as predicted by our
analytical results. In (b) We show plots of $\log(N(s))$ versus $\log (s)$ are
shown for different $k$ values and find straight lines with slopes $-(1-2k)$  as predicted by Eq. (\ref{eq:14}).
} 

\label{fig:3ab}
\end{figure}

Next we solve the rate equations for the case $v(x,t)=k s(t)/t$, where $k$ is a
dimensionless parameter. Here, we  make use of the fact that the mean domain
size of the M-phase decays as $s(t)=1/t$ which is confirmed by numerical simulations 
and found that such a behaviour is independent of precise choice of the growth velocity. Incorporating this
time dependence into the definition {\it (C)} gives  $v(x,t)=k/t^2$. Solving Eq.
(\ref{eq:4}) for this front velocity yields
\begin{equation}
\label{eq:7}
c(x,t)=t^{2(1-k)}\exp[-xt].
\end{equation}
Unlike the previous two cases, this model is of particular interest for the
following reasons. First, all the moments, $M_n(t)$ of $c(x,t)$ where
\begin{equation}
M_n(t)=\int_0^\infty x^nc(x,t)dx,    
\end{equation}
exhibit a power-law   
\begin{equation}
\label{eq:nthmoment}
 M_n(t) \sim t^{-(n-(1-2k))}.  
\end{equation}
Using this relation we find that the mean domain
size $s(t)$ of the M-phase decays as 
\begin{equation}
\label{eq:mean_c}
    s(t)={{M_1(t)}\over{M_0(t)}}=t^{-1}.
\end{equation}
We find that the  mean domain
size decays following exactly the same way as for Model (A) and (B). Secondly, unlike the previous two
cases, the fraction of  the M-phase decays following a power-law
\begin{equation}
\label{eq:8}
\Phi(t)=t^{-2k}.
\end{equation}
To verify this we plot $\log(\Phi(t))$ versus $\log(t)$ for different $k$ values in Fig. (\ref{fig:3a}) using simulation data and 
we find straight lines with slopes equal to $2k$ in each case. It suggests
that our numerical results match perfectly with our analytical solution given by Eq. (\ref{eq:8}).

The physical constraints of the model restrict the feasible range of $k$ values according to Eq. (\ref{eq:nthmoment}).
For instance, the lower bound is fixed by the behaviour of $\Phi(t)$ that must be an increasing function of time 
and hence it demands $k>0$. On the other hand, the upper bound is fixed by the constraint that the 
zeroth moment or the number density should be an increasing function of time or at least during the early and/or at intermediate stage. This immediately provides the upper bound $k<0.5$ since according to Eq. (\ref{eq:nthmoment}) we find 
\begin{equation}
\label{eq:number_c}
    M_0(t)=N(t)\sim t^{1-2k}
\end{equation} 
Thus, the only non-trivial and physically interesting $k$ value are the ones that stay within the interval $[0,0.5]$. One significant result of the present model is that unlike the previous two models here the number density of
domains of M-phase keeps increasing at all time (in the scaling regime). This is due to fact that the front velocity is a decreasing function of time resulting in a decelerated motion of the front and hence if the size of the M-phase is large enough the front will effectively stop at some stage leaving a finite sized interface. In passing, we note that in this model again the average domain size $s(t)$ of M-phase
decays as $s(t)=t^{-1}$. 
The existence of scaling and the non-trivial conservation law provides an
extra motivation to go  beyond the simple scaling description. This can be done
by invoking the idea of fractal analysis, as it has been a very  useful tool to
obtain a global exponent called fractal dimension. To do so, we need a proper
yard-stick to measure the size of the set created in the long time limit. The
most convenient one is the mean domain size $s(t)$. Using  the relation for $s(t)$ we can eliminate $t$ 
from Eq. (\ref{eq:number_c}) and find that the number $N(s)$  scales with yard-stick size or average domain size $s$ 
following a power-law
\begin{equation}
 \label{eq:14}
N(s)\sim s^{-d_f},
\end{equation}
where $d_f=(1-2k)$ provided $k<0.5$. The exponent $d_f$ is known as the fractal dimension or the
Hausdorff-Besicovitch dimension  of the resulting set created
by the nucleation and growth process   \cite{ref.hassan_hassan, kn.fractal,ref.hassan_santo, kn.hassan, ref.fractals_95,ref.hassan_kurths,ref.fractal_self}.
In Fig.  (\ref{fig:3b}) we plot $\log(N(s)))$ versus $\log(s)$ using numerical simulation data for different $k$ and find straight line with slope as it should be according to Eq. (\ref{eq:14}). Once
again it proves that numerical simulation and analytical results match perfectly.

\begin{figure}

\centering

\subfloat[]
{
\includegraphics[height=7.0 cm, width=8.2 cm, clip=true]
{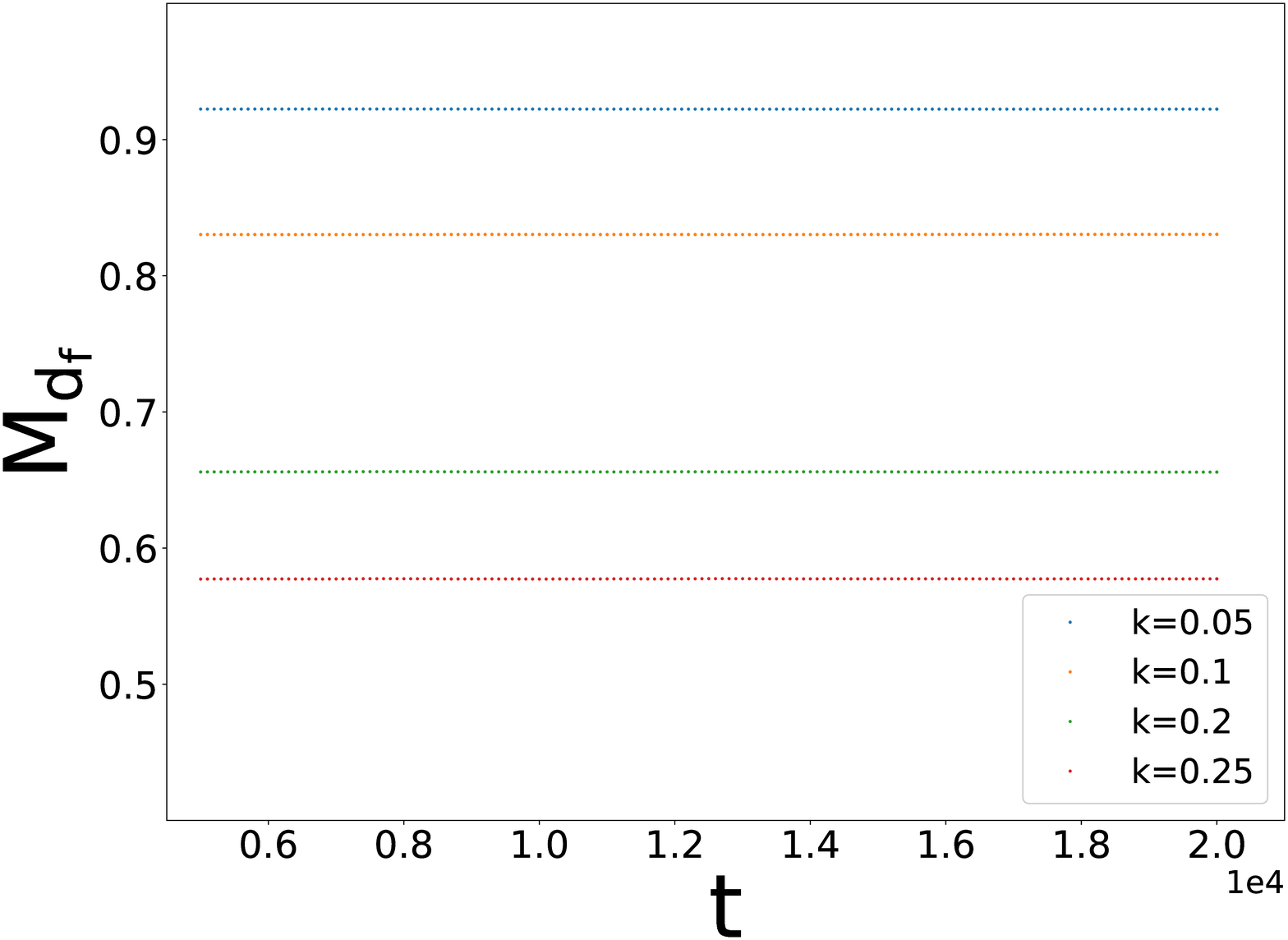}
}

\caption{ Plots of the $d_f$th moment of $c(x,t)$ versus time $t$ are shown for different values of $k<0.5$. It clearly 
demonstrates that the $d_f$th moment is always a conserved quantity suggesting
that numerical and analytical results are in perfect agreement. 
} 

\label{fig:4}
\end{figure}

Despite the numerical values of the major variables of the system, interval sizes $x_i$, are changing with time yet according to Eq. (\ref{eq:nthmoment})
we observe that $(1-2k)$th moment of $c(x,t)$ is a conserved quantity. That is,
at any given time if there are $m$ number of intervals of M-phase of size
$x_1,x_2,\cdots, x_m$ then 
\begin{equation}
 M_{1-2k}(t)=x_1^{1-2k}+x_2^{1-2k}+\cdots + x_m^{1-2k}={\rm const.},   
\end{equation}
regardless of the time and the value of $m$ provided $k<0.5$ (see Fig. (\ref{fig:4})). 
Note that self-similarity along the continuous time axis is a kind of symmetry.
Thus, on one hand, we find that the system enjoys dynamical scaling symmetry that
manifests itself through data collapse and on the other, we have the $(1-2k)$th moment 
of the interval size of the M-phase is a conserved quantity in time. 
The emergence of non-trivial conserved quantity accompanied by continuous self-similar symmetry in time 
is reminiscent of Noether's theorem as it states that for every symmetry there is a corresponding conservation 
law and vice versa \cite{ref.fractal_conserved, ref.fractal_rakib}.

\begin{figure}

\centering

\subfloat[]
{
\includegraphics[height=4.0 cm, width=4.2 cm, clip=true]
{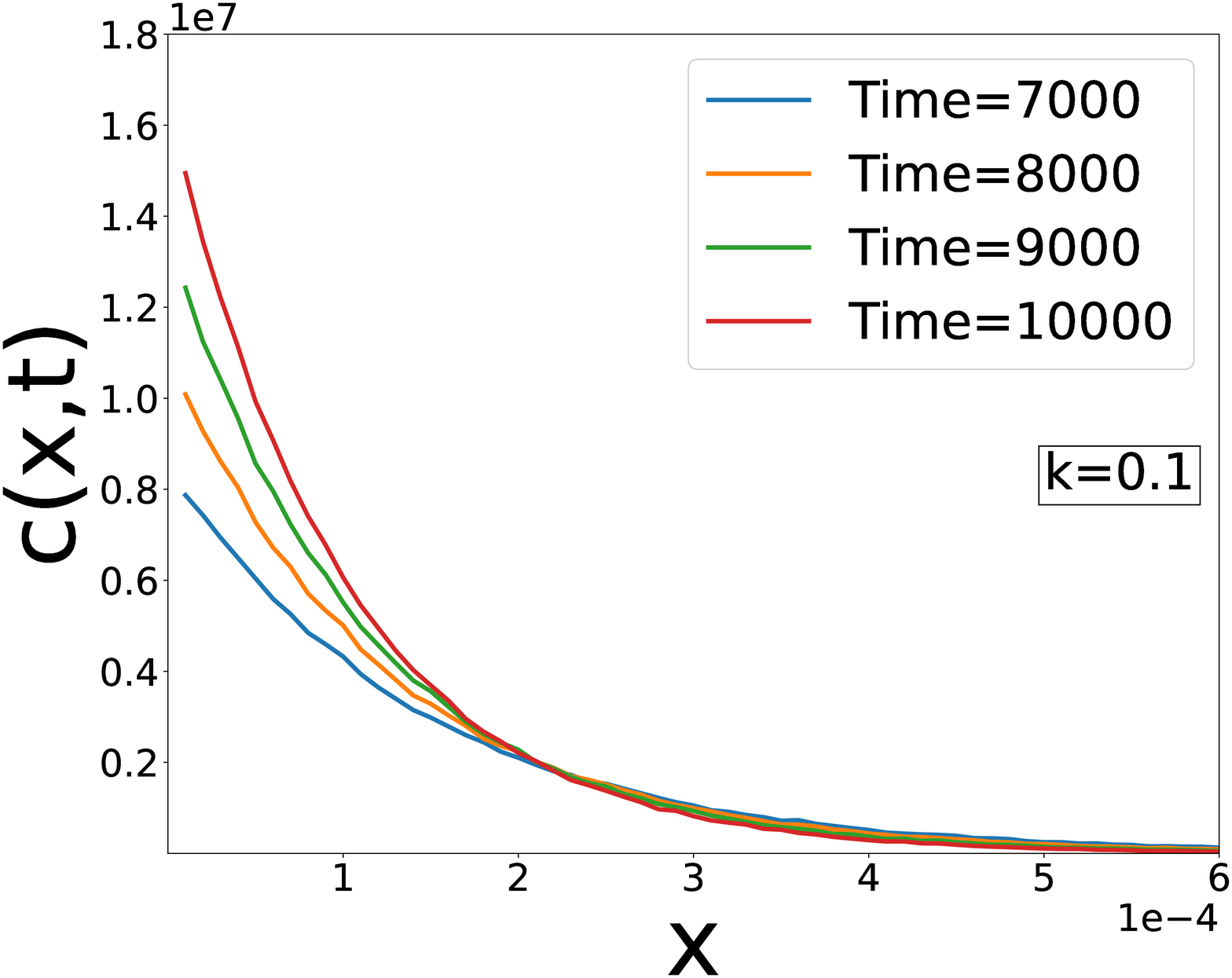}
\label{fig:5a}
}
\subfloat[]
{
\includegraphics[height=4.0 cm, width=4.2 cm, clip=true]
{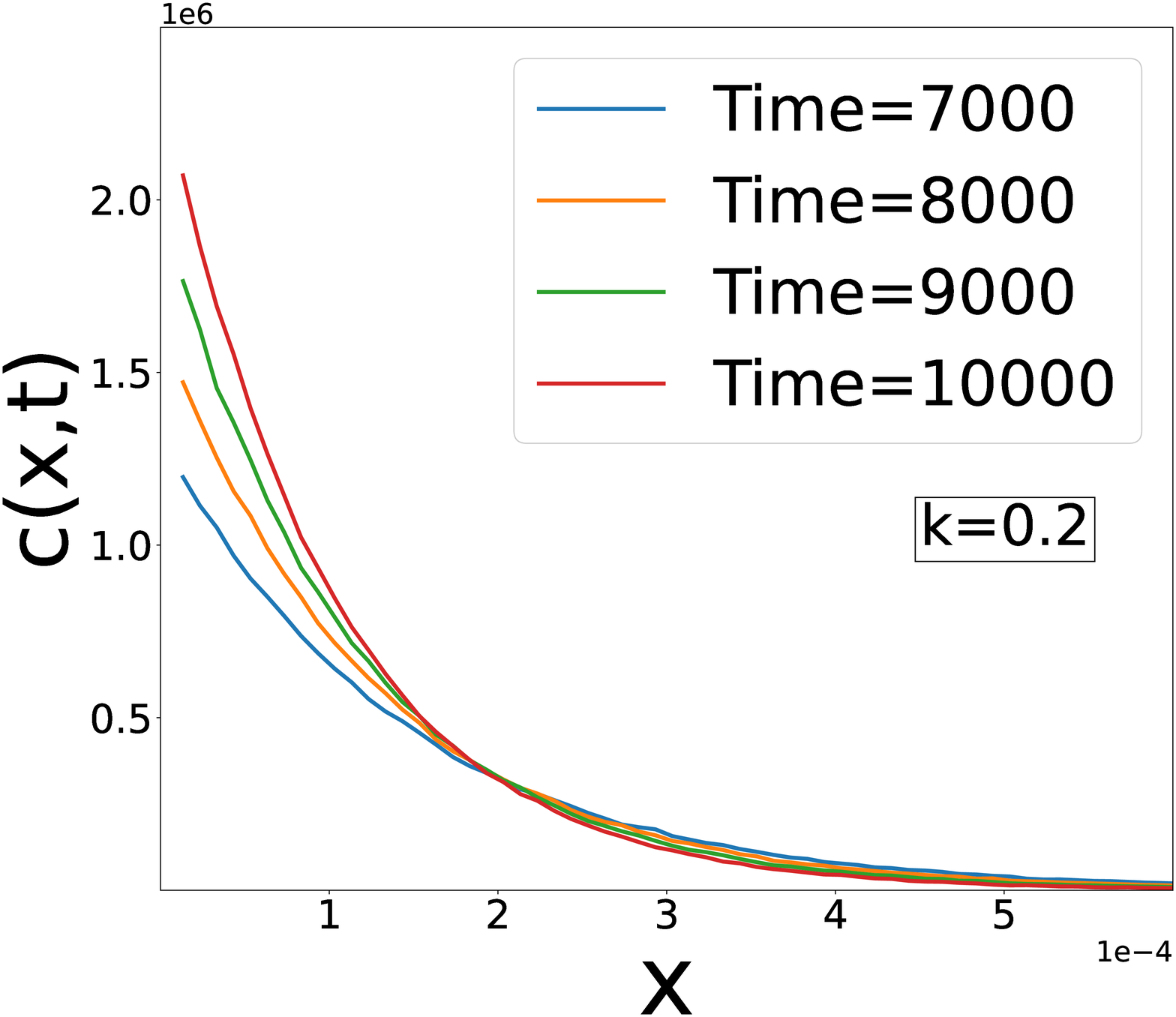}
\label{fig:5b}
}

\subfloat[]
{
\includegraphics[height=4.0 cm, width=4.2 cm, clip=true]
{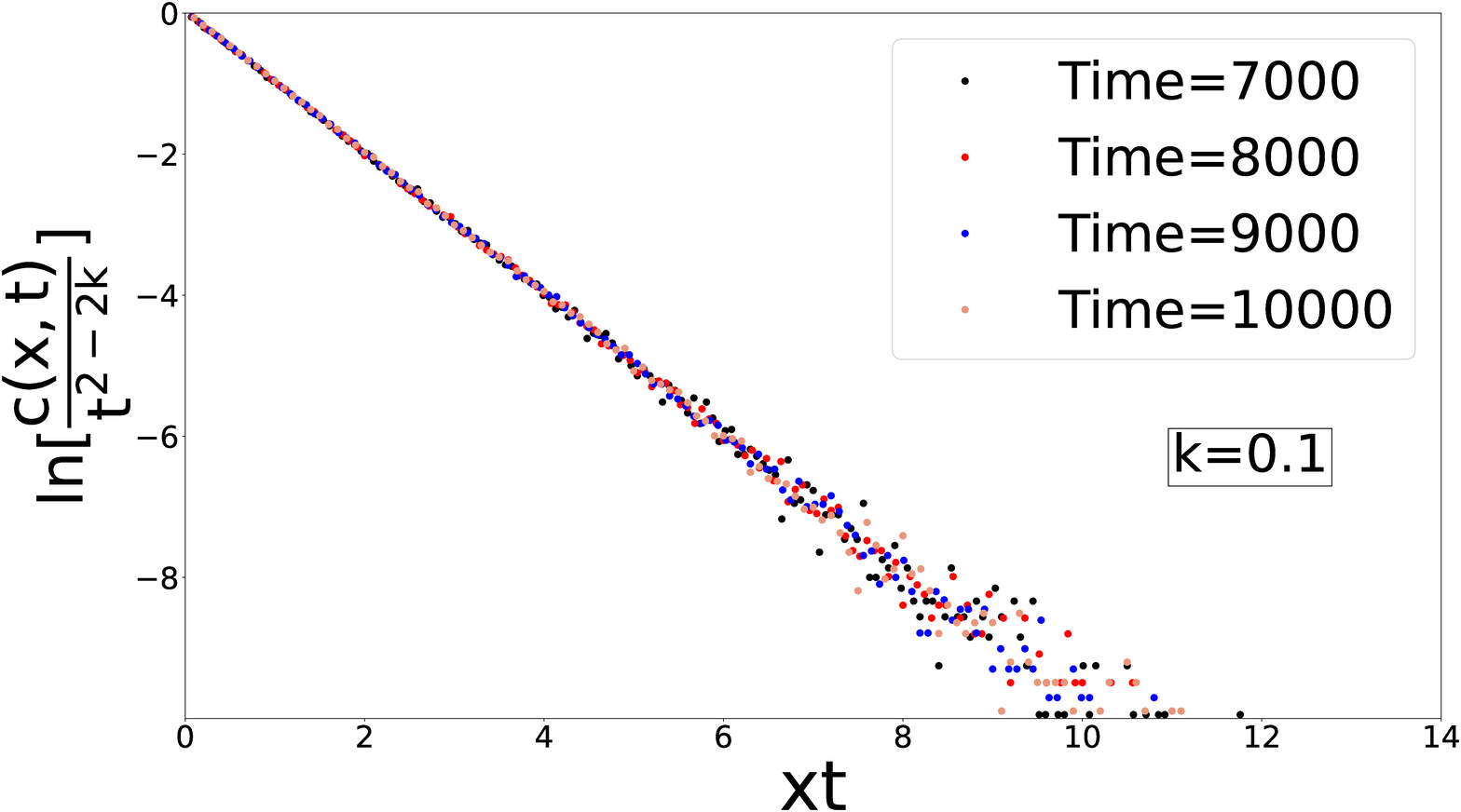}
\label{fig:5c}
}
\subfloat[]
{
\includegraphics[height=4.0 cm, width=4.2 cm, clip=true]
{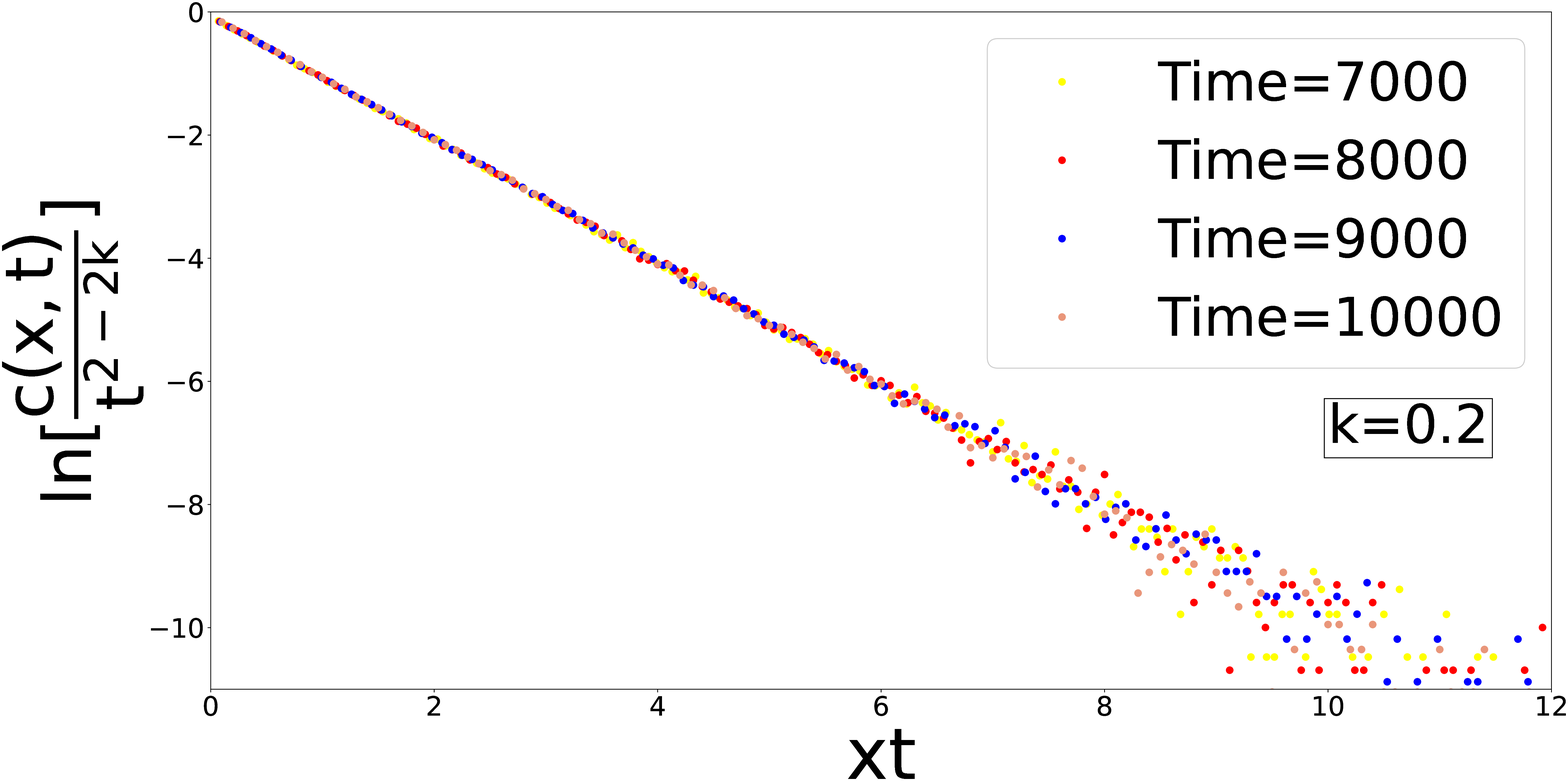}
\label{fig:5d}
}

\caption{ The plots of the distribution function $c(x,t)$ vs $x$ are shown  in (a) and (b)  
for $k=0.1$ and $k=0.2$ respectively. Each plot represents four different times where each set of data represents ensemble average over
$1000$ independent realizations.
In (c) and (d) we plot $\log[c(x,t)/t^{2-2k}$ vs $xt$ using the data of (a) and (b) respectively and find 
that all the data collapse into a straight line with slope equal to $-1$ as suggested by Eq. (\ref{eq:7}).  
} 

\label{fig:5abcd}
\end{figure}

We know that if a function $f(x,t)$ of a time varying phenomena satisfies the condition
\begin{equation}
\label{eq:dynamic_scaling}
    f(x,t)\sim t^\theta\phi(x/t^z),
\end{equation}
then it is said to obey dynamic scaling \cite{ref.hassan_ba_dc,ref.hassan_sarkar_liana_dc,ref.fractal_cda}. The solution for $c(x,t)$ given by Eq. (\ref{eq:7}) has exactly
the same form as in Eq. (\ref{eq:dynamic_scaling}) provided we choose $\theta=2(1-k)$
and $z=-1$. Here, the exponent $\theta$ and $z$ are fixed by the dimensional requirement that
$[f]=[t^{\theta }]$ and $[x]=t^z]$ respectively. Thus, the model (c) exhibits dynamic scaling
with $\theta=2(1-k)$ and $z=-1$. It implies that the numerical value of $c(x,t)$
varies with $x$ for a given time $t$. That is, the data of $c(x,t)$ versus $x$ will be distinct 
for fixed time and for each different $k$ value which can be seen in Figs. (\ref{fig:5a}) and  (\ref{fig:5b}). 
However,  if we plot $c(x,t)t^{-2(1-k)}$ or  $c(x,t)t^{-(1+d_f)}$ versus $xt$ they should collapse
into one universal scaling curve. Indeed, they do so, however to show that the scaling function $\phi(\xi)$ is exponential
we plot them in the $\log$-linear scaling and find that all the distinct curves  in 
Figs. (\ref{fig:5a}) and  (\ref{fig:5b}) collapse superbly onto their respective universal curves which are 
shown respectively in Figs. (\ref{fig:5c}) and  (\ref{fig:5d}). It means that the solution for the scaling
function is exponential. Besides, the data collapse means that the system evolves with time and that the snapshots 
taken at different times are similar. Since the same system at different times are similar the solutions are self-similar.
Self-similarity in this problem manifests, only statistically, through
dynamical scaling. This is one of the key properties of fractal too.

To verify these results we have done Monte Carlo simulation based on the
algorithm described in Model A except step (iii) which is replaced as follows. 
\begin{itemize}
\item Increase the size of the seed of S-phase on either side by $k/j^2$
independently provided domain of S-phase is greater than $\sigma$ in both sides. However,
if the domain size of S-phase in any side is less than $\sigma/j^2$ then the growth ceases immediately at the point of contact, while it continues elsewhere. 
\end{itemize}
The data for all the plots shown for this model are obtained by Monte Carlo simulation based on the algorithm described above. The
perfect matching with our analytical solution suggests that the algorithm is ideal for describing the integro-differential
equation given by Eq. (\ref{eq:RateEq}) with growth velocity $v=k/t^2$.

\section{Model D} \label{sec-model_d}

\begin{figure}

\centering

\subfloat[]
{
\includegraphics[height=4.0 cm, width=4.2 cm, clip=true]
{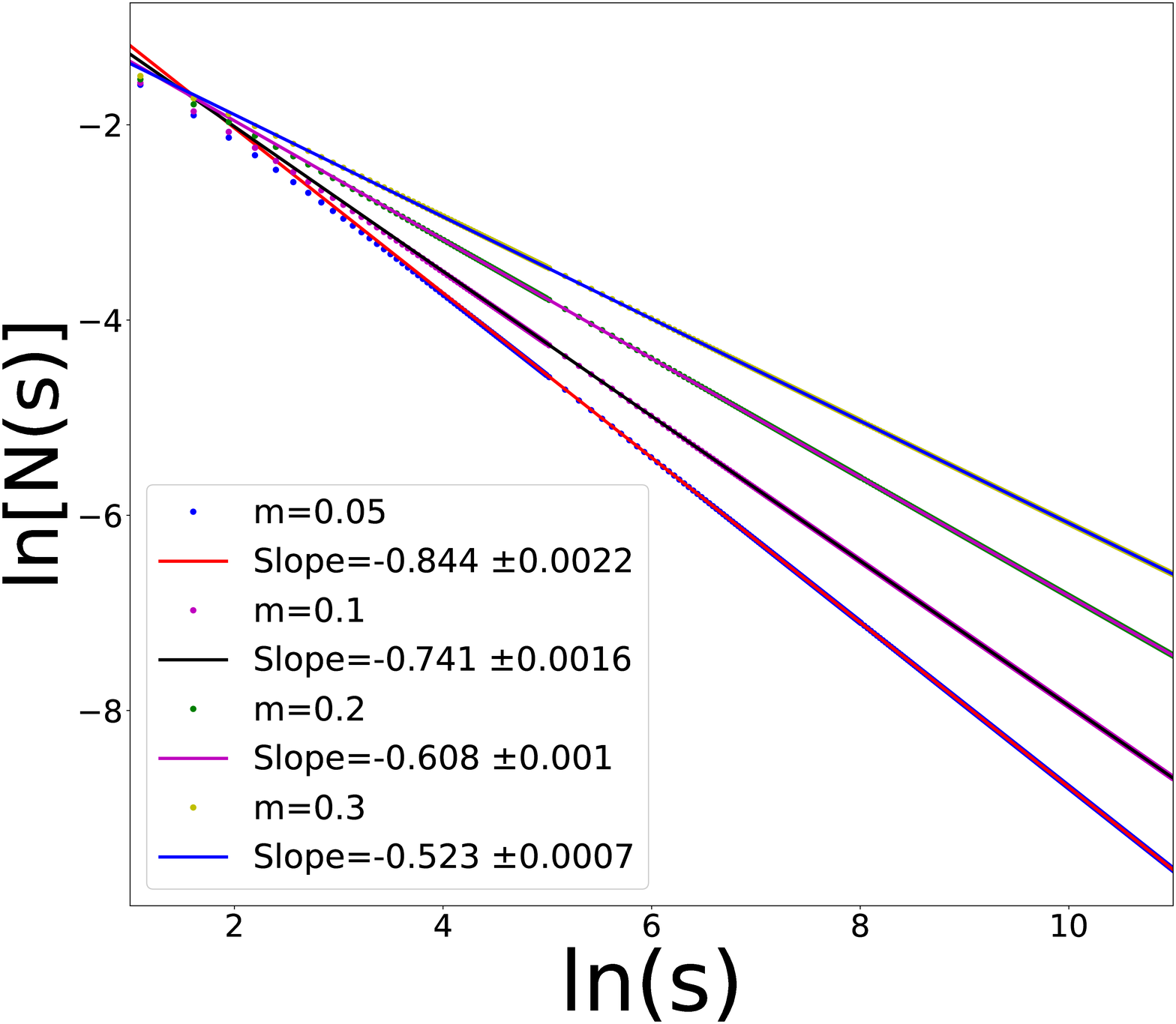}
\label{fig:6a}
}
\subfloat[]
{
\includegraphics[height=4.0 cm, width=4.2 cm, clip=true]
{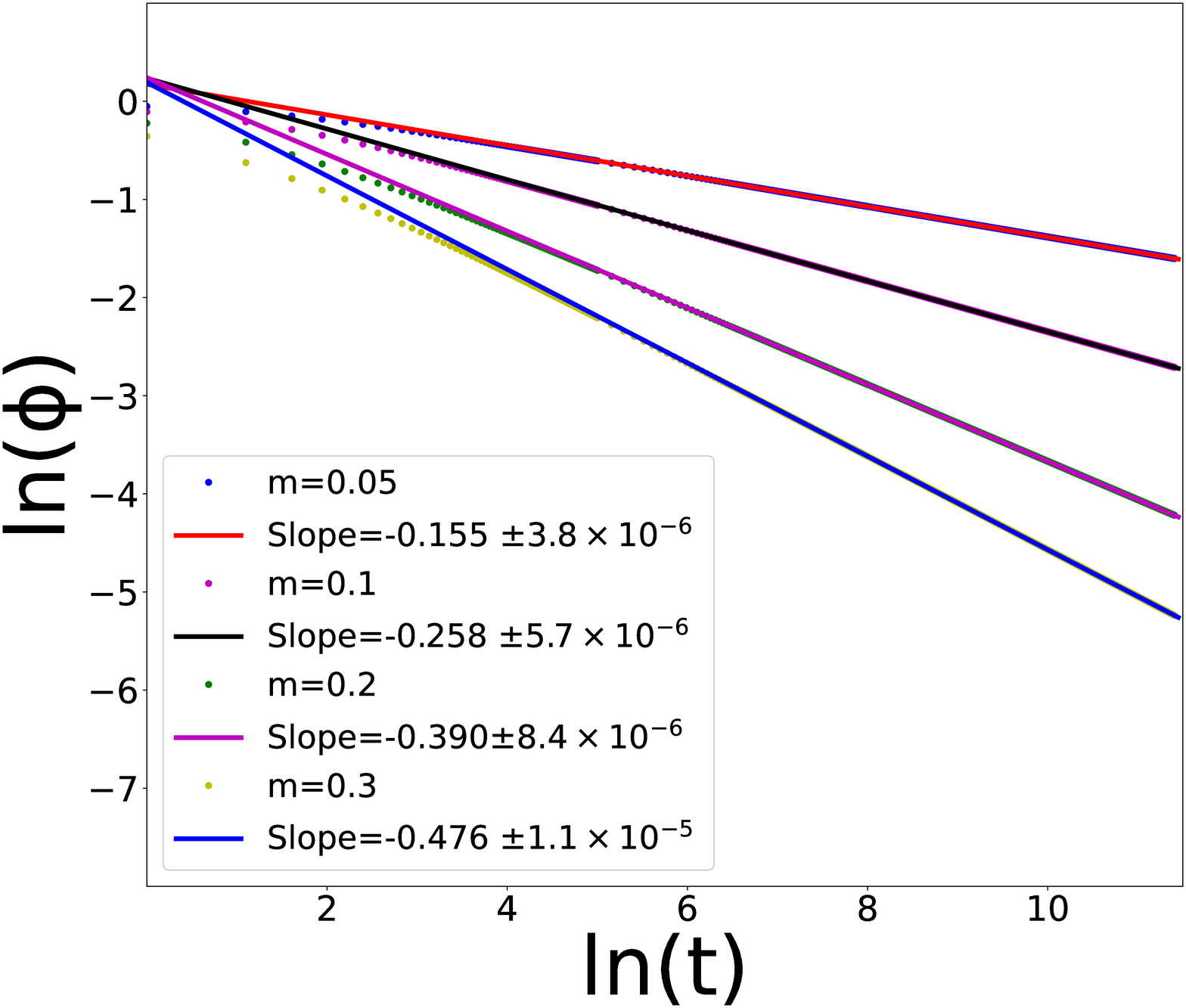}
\label{fig:6b}
}
\caption{Plots of $\log(N(s))$ versus $\log (s)$, where $N$ is number of domains of M-phase, are
shown in (a) for different $m$ values. In (b) we show plots of $\log(\Phi(t))$ versus
$\log(t)$ for different $m$ values. In each case it results in a straight line with slopes
always equal to $-d_f$ in (a) and $-(1-d_f)$ in (b). These results match perfectly with our
analytical results.
}

\label{fig:6ab}
\end{figure}

\begin{figure}

\centering

\subfloat[]
{
\includegraphics[height=7.0 cm, width=8.2 cm, clip=true]
{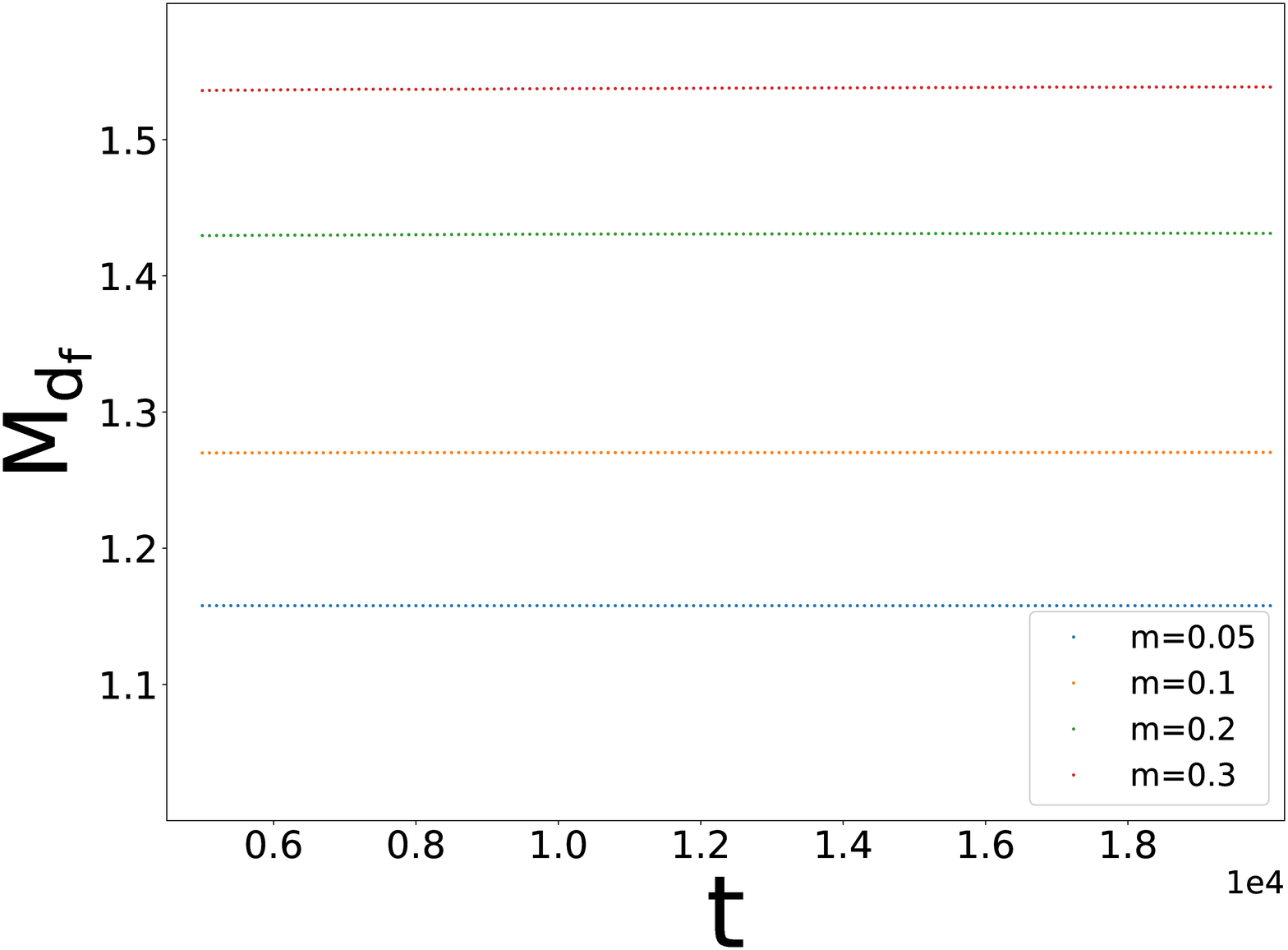}
}

\caption{ Here we show that the $d_f$th moment of $c(x,t)$ is always a conserved quantity for all $m$ values of model D.
} 

\label{fig:7}
\end{figure}

Finally, we consider the case where $v(x,t)= mx/\tau$ which implies that the
S-phase travels a distance $x$ of the M-phase in time $\tau$ transforming $x$ 
into the S-phase \cite{kn.maslov,ref.fractal_cda}. However, according to Eq.(\ref{eq:RateEq}), the typical time between
two nucleation events on an interval  of size $x$ is $\tau=x^{-1}$ and therefore
$v(x,t)=mx^2$ revealing that the growth velocity  decreases increasingly fast as
the domain size of the M-phase decreases since $x$ itself is a decreasing quantity with
time.  To solve the model we  incorporate the
definition of $M_n(t)$ into Eq. (\ref{eq:RateEq}) and then, after some simple
algebraic  manipulations, we are able to write the following rate equation for
$M_n(t)$
\begin{equation}
\label{eq:12}
{{dM_n(t)}\over{dt}}=-\Big [{{n^2+n(1+{{1}\over{2m}})-{{1}\over{2m}}}\over{(n+1)}}\Big ]M_{n+1}(t).
\end{equation}   
Note that for $m=0$ the total mass or $M_1(t)$ is a conserved quantity.
However, for $m>0$ the system violates this  conservation of mass principle due to continuous
growth of $S$-phase at the expense of $M$-phase. The interesting feature of the above equation
is that for each value of $m$ there exists a unique conserved quantity. We can find the
value $n=\gamma$ for which the moment $M_\gamma(t)$ is a conserved quantity and it is
done simply by setting $dM_n(t)/dt = 0$ \cite{ref.krapivsky_naim_fractal}.
Note that $dM_n(t)/dt = 0$  can be equal to zero either due to $M_{n+1}(t)=0$ of Eq. (\ref{eq:12}) or due to its co-factor equal to zero. 
If the former is zero then it leads to trivial result. On the other hand, if the latter is true then the 
problem rests on solving a quadratic equation in $\gamma$ whose real positive root is
\begin{equation}
\label{eq:m}
\gamma(m)=-{{1}\over{2}}\Big (1+{{1}\over{2m}}\Big )+{{1}\over{2}}\sqrt{\Big (1+{{1}\over{2m}}\Big )^2+{{2}\over{m}}}.
\end{equation}
It implies that the corresponding moment $M_{\gamma(m)}(t)$ is independent of time. It has been verified by
numerical simulation as shown in Fig. (\ref{fig:7}). To find the significance of the value
of $\gamma$ we once again invoke the idea of fractal.  Like in model (C), we can use $s(t)$ as a yard-stick and obtain the $N(s)$ we need to cover the system. 

We can solve Eq. (\ref{eq:12}) by assuming a trial solution 
\begin{equation}
    M_n(t)\sim t^{\alpha(n)},
\end{equation}
such that $\alpha(\gamma)=0$ since $M_{\gamma}(t)$ is a conserved quantity. Substituting it in Eq. (\ref{eq:12}) and demanding dimensional consistency we find a recurrence relation
\begin{equation}
    \alpha(n+1)=\alpha(n)-1.
\end{equation}
Iterating it over and over again subject to the condition that $\alpha(\gamma)=0$ as required by the conservation law
we find that
\begin{equation}
    M_n(t)\sim t^{-(n-\gamma(m))}.
\end{equation}
Using this we find that the average domains of the M-phase $s(t)=M_1(t)/M_0(t)$
decays exactly like all previous three cases i.e., $s(t) = t^{-1}$. 
We
then find that $N(s)$ scales as
\begin{equation}
 \label{eq:14d}
N(s)\sim s^{-d_f},
\end{equation}
where $d_f(m)$ is given by Eq. (\ref{eq:m}). The plots of $\log(N(s))$ versus $\log(s)$ for different
$m$ are shown in Fig. (\ref{fig:6a}) using data from extensive Monte Carlo simulation and find excellent straight
lines with slopes equal to $d_f$ \cite{ref.hassan_hassan,ref.fractal_self}.
The exponent $d_f$ is known as the fractal dimension or the
Hausdorff-Besicovitch dimension   of the resulting set created
by the nucleation and growth process \cite{kn.fractal}. Besides, like in model (C),  the Kolmogorov-Avrami formula in this
case is no longer valid. Instead, it is  replaced by the following general
power-law decay of the M-phase
\begin{equation}
\label{eq:15}
\Phi(t)\sim t^{-(1-d_f)}.
\end{equation}
In Fig. (\ref{fig:6b}) we show that plots of $\log(\Phi(t)$ versus $\log(t)$ for different $m$ and find that they all
obey Eq. (\ref{eq:15}). This reveals a generalised exponent $(1-d_f)$ that can quantify the extent of the decay of 
the M-phase. On the other hand, the coverage by the S-phase $\theta(t)$ reaches its asymptotic value 
$\theta(\infty)=1$ as 
\begin{equation}
\theta(\infty)-\theta(t) \sim t^{-(1-d_f)},
\end{equation}
which is reminiscent of the Feder's law in RSA processes \cite{kn.fractal}.

\begin{figure}

\centering

\subfloat[]
{
\includegraphics[height=4.0 cm, width=4.2 cm, clip=true]
{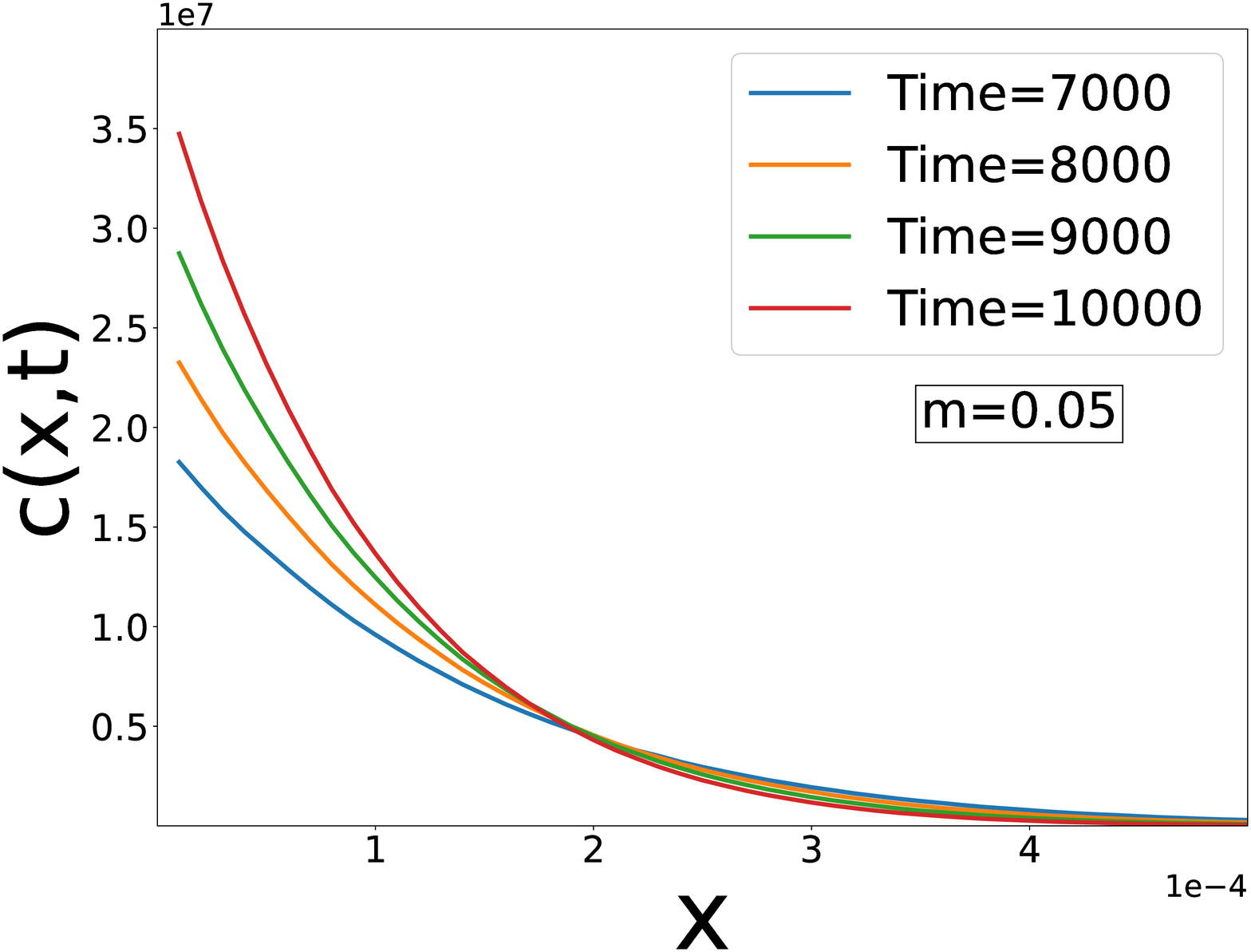}
\label{fig:8a}
}
\subfloat[]
{
\includegraphics[height=4.0 cm, width=4.2 cm, clip=true]
{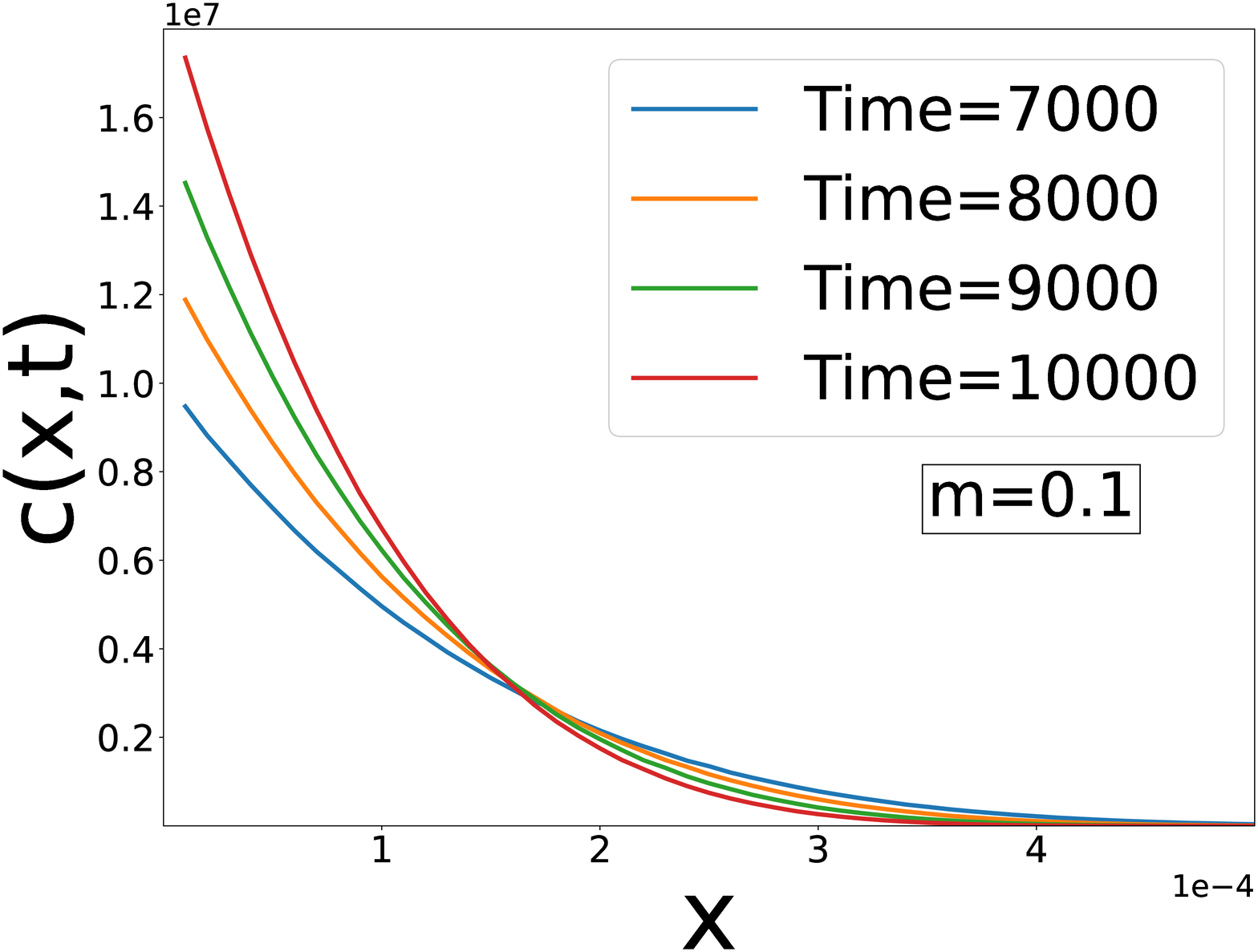}
\label{fig:8b}
}

\subfloat[]
{
\includegraphics[height=4.0 cm, width=4.2 cm, clip=true]
{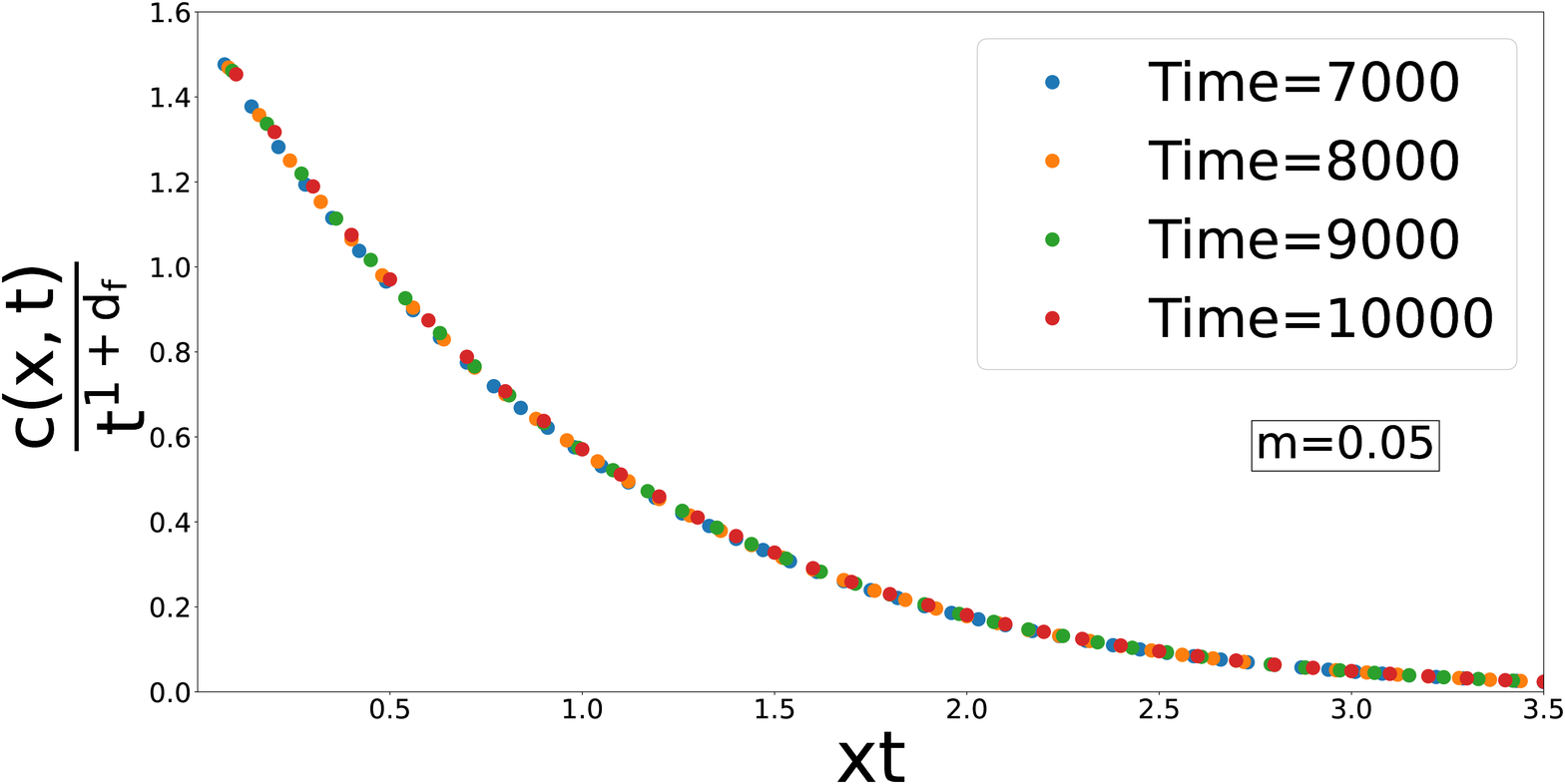}
\label{fig:8c}
}
\subfloat[]
{
\includegraphics[height=4.0 cm, width=4.2 cm, clip=true]
{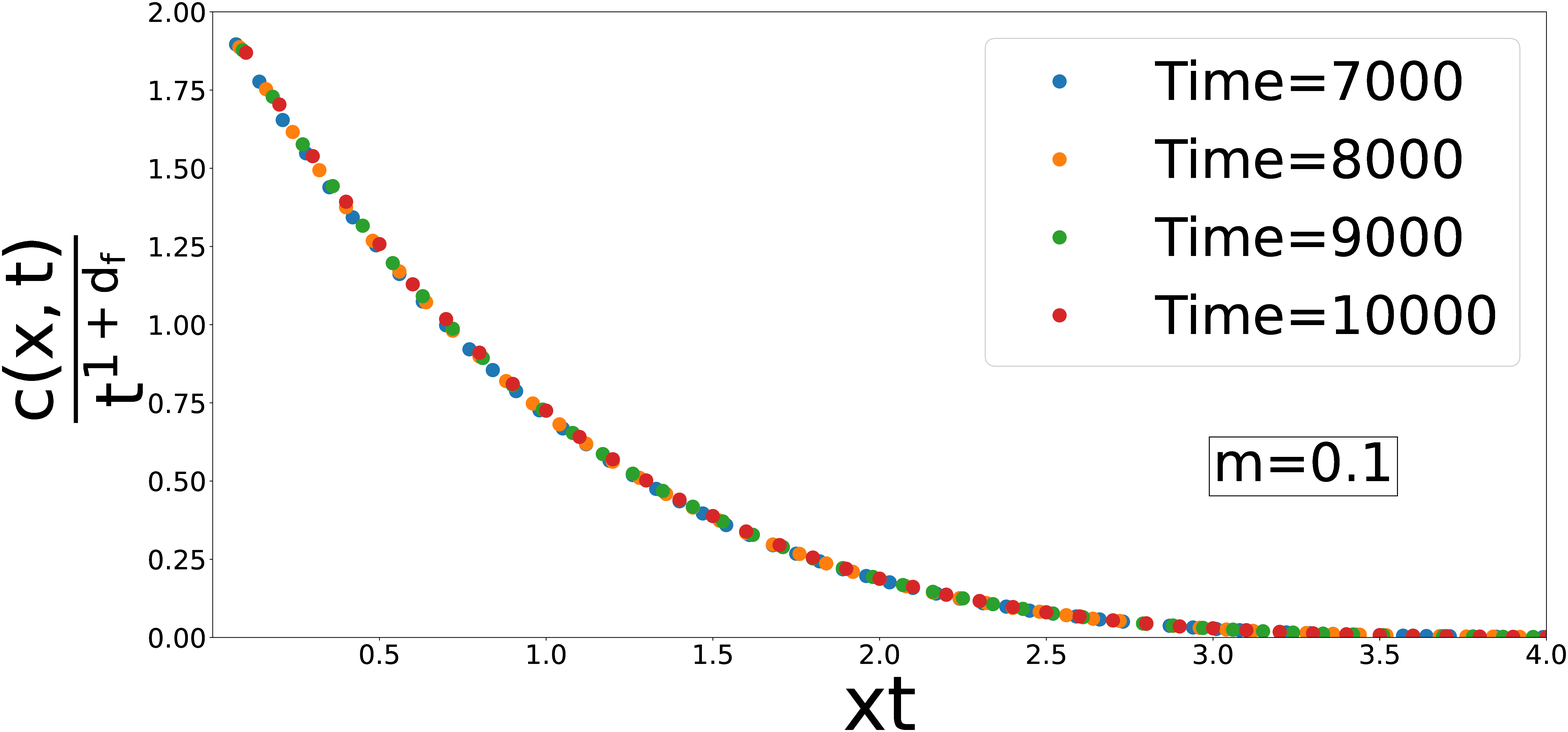}
\label{fig:8d}
}

\caption{Plots of $c(x,t)$ versus $x$ for (a) $m=0.05$ and (b) $m=0.1$ are drawn as a representative
values of $m$. In (c) and (d) we plot $c(x,t)t^{-(1+d_f)}$ versus $xt$ and we find excellent data collapse of the
same data of (a) and (b) respectively which confirms that the model D obeys dynamic scaling.
} 

\label{fig:8abcd}
\end{figure}

Finally, we show that the solution for $C(x,t)$ exhibits dynamic scaling. To do this, we check 
if the solution for $c(x,t)$ obeys dynamic scaling 
\begin{equation}
 \label{eq:13}
c(x,t)\sim t^{(1+d_f)}\phi(xt).
\end{equation} 
To verify this we plot $c(x,t)$ versus $x$ in Figs. (\ref{fig:8a}) and (\ref{fig:8b}) for different
$m$ value. According to Eq. (\ref{eq:13}) the same data would collapse if we divide $c(x,t)$ by
$t^{\sqrt{2}}$ and $x$ by $t^{-1}$. Indeed, we find excellent data collapse as shown in  Figs. (\ref{fig:8c}) 
and (\ref{fig:8d}) respectively for $m=0.05$ and $m=0.1$. It provides a clear litmus test of our solution 
that it obeys
dynamic scaling and hence like model C, the snapshots of the model D too taken at different times are similar. 
It is worth to mention that the number density in {\it (C)} and {\it
(D)} increases for all  time - a sharp contrast to the models {\it (A)} and {\it
(B)} where  $N(t)$ increases only at the early stage. This is due to the fact
that in models {\it (C)} and {\it (D)},  the growth velocity decreases in time
in such a way that two growing phases from opposite direction  hardly coalesce.
In fact, the growth of the S-phase virtually stops at some stage.

To verify these results we have done Monte Carlo simulation based on the
algorithm described in Model A except step (iii) which is replaced as follows. 
\begin{itemize}
\item Increase the size of the seed of S-phase on each side by an amount equal to
square of the respective domain size of the M-phase and the growth ceases immediately at the point of contact, while it continues elsewhere. 
\end{itemize}

\begin{table}
\begin{tabular}{|l|l|l|l|l|l|l} \hline
Model & $\Phi(t)$ & $N(t)$ & $s(t)$ & Scaling & Fractal \\ \hline
$v=v_0$ & $e^{-v_0t^2}$ & $ t\, e^{-v_0t^2}$ & $t^{-1}$ & violates & $\times $ \\ \hline
$v=\sigma/t$ & $e^{-2\sigma t}$ & $t\,e^{-2\sigma t}$ & $t^{-1}$ & violates & $\times $ \\ \hline
$v=k s(t)/t$ &  $t^{-(1-d_f)}$ & $t^{d_f}$ & $t^{-1}$ & obeys & $(1-2k)$ \\ \hline
$v=m x^2$ & $t^{-(1-d_f)}$ & $t^{d_f}$ & $t^{-1}$ & obeys & $d_f(m)$ \\ \hline
\end{tabular}
\caption{Summary of the various model results}
\end{table}

\section{Discussion of findings and summary} \label{sec-conc}

In this article, we have studied a class of random nucleation and growth processes of a
stable phase with three different growth velocities which can be space and time dependent.
Also, we revisit the classical KJMA model. The three new growth velocities describe a process in which
the growth velocity depends on the local configuration of the system at every
instant of time. We have solved the models analytically and verified the results 
by extensive Monte Carlo simulation. One interesting finding is that in all four cases 
the mean domain size $s(t)$ of M-phase decays following  a
power-law $s(t)\sim t^{-\alpha}$ with the same exponent $\alpha=1$ irrespective of the detailed choice of the
growth velocity. There must be a common mechanism behind this common nature. Indeed, all the four models
share one common thing, i.e. the nucleation process is described by the random
secession model. However, the fraction of the untransformed space
(M-phase) and the corresponding number density are very sensitive to the specific choice of
the growth velocity. 

The most striking result, though, is the emergence of fractal
for either fully time  dependent or fully size dependent growth velocities in the sense that the growth velocities 
are either proportional to inverse square of time $t$ or simply proportional to the square of size $x$.
The system is called fractal if the exponent of the power-law decay of $N(s)$ as a function of $s$ is less
than the dimension $d=1$ of the space where the system is embedded and at the same time the system must be self-similar. 
One of the ways to test whether the systems that evolves
probabilistically with time is self-similar or not is that it must exhibit dynamic scaling. 
Testing of dynamic scaling means that the numerical values of the dimensional quantities such as $c(x,t)$ and $x$ 
at different time will be different but the corresponding dimensionless quantities $c(x,t)/t^{\theta z}$ and $x/t^z$ would coincide. In other words, the plots
of $c(x,t)$ versus $x$ for time will be distinct but all these distinct plots would collapse if we plot $c(x,t)/t^{\theta z}$ versus $x/t^z$ instead. We
find $\theta=1+d_f$ and $z=-1$ for both model C and D albeit the fractal dimension $d_f$ is different. 
The self-similarity is also a kind of continuous symmetry along the
continuous time axis. Interestingly, we also find that during the evolution the $d_f$th moment is always a conserved quantity. These two results are reminiscent of Noether's
theorem that states that for every continuous symmetry there must exist a conserved quantity. We have also shown that when either the temporal
variable $t$ or both spatial and the temporal in the definition of the velocity are assumed constant, the decay of the metastable phase is always exponential
and is also accompanied by the violation of scaling. 
One of the key observables in nucleation and growth processes is how the fraction of the M-phase decays to let the new S-phase grow. We find
that both in model C and D, it decays following a power-law with a unique generalised exponent $(1-d_f)$ though the $d_f$ values are different.

Thus, the present work represents the most obvious natural extension of
the classical KJMA theory. We believe that this extension is a realistic
description for many natural processes. Furthermore, our model sketches
the process of invasion and growth of a biological colony with a random
immigration of pioneer individuals or seed arrival. One important example is the
spread of diseases such as fungal spots on plant leaves. Here one assumes that
spores are transmitted by wind or insects and randomly arrive on the leave
surface. Once an infected spot is established it stays and begins to grow. In
many studies of such biological processes the growth has been found to deviate
from the constant velocity assumption of the KJMA model. Instead it is typically
observed that the velocity increases with the increase in proportion of non-infected areas,
i.e. the M-phase (Vanderplank principle  \cite{Vander}). Therefore we
believe that the present investigation of the consequences of state-dependent
growth velocities will have important applications.

We conclude with the following words. The power-law decay of the M-phase can be
assumed to be a generalized formula replacing the classical Kolmogorov-Avrami
law, provided, the distribution of the M-phase in the late stage describes a
scale-free fractal. The fractal dimension is the quantitative measure of the
notion that the density of the M-phase is less at larger length scale. We
believe that the present work will have a significant impact in changing the way
we  intended to interpret the experimental data as we are now aware of the
fascinating results due to the decelerating growth velocity.

\end{document}